\documentclass[11pt]{article}
\pdfoutput=1  
\usepackage{jcappub}
\usepackage{graphicx}
\usepackage{amsmath,amssymb}
\graphicspath{{./figures/}}
\usepackage{appendix}
\usepackage{placeins}
\usepackage[normalem]{ulem}
\usepackage[utf8]{inputenc}

\input epsf
\usepackage{mathtools}
\usepackage{amsfonts}
\usepackage{upgreek}
\usepackage{latexsym}
\usepackage{enumitem}
\usepackage{stfloats}
\usepackage{afterpage}

\newcommand{\be}{\begin{equation}}
\newcommand{\ee}{\end{equation}}
\newcommand{\beq}{\begin{equation}}
\newcommand{\eeq}{\end{equation}}
\newcommand{\bea}{\begin{eqnarray}}
\newcommand{\eea}{\end{eqnarray}}

\newcommand{\bx}{{\mathbf  x}}
\newcommand{\by}{{\mathbf  y}}

\newcommand{\bn}{{\mathbf  n}}
\newcommand{\bk}{{\mathbf k}}
\newcommand{\boe}{{\mathbf  e}}

\newcommand{\bV}{{\mathbf V}}

\newcommand{\bal}{\boldsymbol{\al}}
\newcommand{\bell}{\boldsymbol{\ell}}

\newcommand{\dd}{\partial}
\newcommand{\HH}{{\cal H}}

\newcommand{\De}{\Delta}

\newcommand{\PP}{{\cal P}}

\newcommand{\RR}{{\cal R}}

\newcommand{\al}{\alpha}

\newcommand{\ga}{\gamma}

\newcommand{\de}{\delta}
\newcommand{\ep}{\epsilon}
\newcommand{\ka}{\kappa}

\newcommand{\La}{\Lambda}

\newcommand{\Om}{\Omega}

\newcommand{\Stwo}{{\mathbb{S}^2}}
\newcommand{\cd}{\cdot}
\newcommand{\ra}{\rightarrow}
\newcommand{\pa}{\parallel}
\newcommand{\class}{{\sc class}}

 \definecolor{magenta}{rgb}{0.1,0.98,0.6}

\definecolor{drot}{rgb}{0.6,0.0,0.0}
\definecolor{dgreen}{rgb}{0.0,0.60,0.0}

\definecolor{cyan}{rgb}{0,0.6,0.6}

 \definecolor{darkgreen}{rgb}{0.2, 0.7, 0.23 }

\title{The Flat Sky Approximation to Galaxy Number Counts}
\author{William L. Matthewson}
\author{and Ruth Durrer}
\affiliation{Universit\'e de Gen\`eve, D\'epartement de Physique Th\'eorique and Centre for Astroparticle Physics,
24 quai Ernest-Ansermet, CH-1211 Gen\`eve 4, Switzerland}

\emailAdd{william.matthewson@unige.ch}
\emailAdd{ruth.durrer@unige.ch}

\abstract{
We derive and test an approximation for the angular power spectrum of galaxy number counts in the flat sky limit. The standard density and redshift space distortion (RSD) terms in the resulting approximation are distinct to the Limber approximation, providing an accurate result for multipoles as low as $\ell\simeq10$, where the corresponding Limber approximation is completely inaccurate.  At equal redshift the accuracy of the density and RSD (standard) terms is around 0.2\% for $z<3$ and 0.5\% at $z=5$,  even to $\ell<50$. 
At unequal redshifts, if we consider the total power spectrum, the precision is better than 5\% only for very small redshift differences, $\delta <\delta_0 (\simeq 3.6\times10^{-4}(1+z)^{2.14})$ where the standard terms are well-approximated, or for large enough redshift differences $\delta >\delta_1 (\simeq 0.33(r(z)H(z))/(z+1))$ where the lensing terms dominate. The flat sky expressions for the pure lensing and the lensing-density cross-correlation terms are equivalent to the Limber approximation. For arbitrary redshift differences, the Limber approximation achieves an accuracy of 0.5\% (above $\ell\simeq 40$ for pure lensing and $\ell\simeq 80$ for density-lensing).  Besides being very accurate, the flat sky approximation is computationally much simpler and can therefore be very useful for data analysis and forecasts with MCMC methods. This will be particularly crucial for upcoming galaxy surveys that will measure the power spectrum of galaxy number counts.
}

\begin{document}

\maketitle

\section{Introduction}
At present, the most successful cosmological data are the anisotropies and polarisation of the cosmic microwave background (CMB), see~\cite{Aghanim:2018eyx} for the latest experimental results. However, in this decade there are deep and large galaxy surveys planned~\cite{Abell:2009aa,Abate:2012za,Laureijs:2011gra,Amendola:2016saw,Blanchard:2019oqi,Aghamousa:2016zmz,Dore:2014cca,Maartens:2015mra,Santos:2015gra} which may do as well as, or in some aspects better than, CMB experiments. To make optimal use of these data, a correct analysis has to be performed. On small scales and at late times, non-linearities and baryonic effects are the most difficult challenge, while at intermediate to large scales and higher redshifts a correct relativistic treatment is most relevant.

The power spectrum of galaxy number counts that upcoming galaxy surveys promise to provide will contain a wealth of cosmological information. We will be able to constrain various cosmological parameters to a level that we can motivate or rule out different dark energy models, measure the neutrino mass hierarchy, learn about inflation e.g. via non-Gaussianity and test General Relativity on cosmological scales, see e.g~\cite{Maartens:2015mra,Amendola:2016saw}. In order to perform an MCMC likelihood analysis, the power spectrum has to be calculated for about $10^5$ to $10^6$ different values for the cosmological parameters, and so it is imperative that spectrum calculations  take as little time as possible. Existing codes such as {\sc CLASS} or {\sc CAMB sources} may be used to perform the full calculation, or to calculate the galaxy number counts power spectrum using the Limber approximation. However, as we shall see, the Limber approximation is very in-accurate for density and redshift space distortions. Therefore it is useful to investigate other approximations that involve simpler calculations in the hope that they might be accurate enough. In this paper, we focus on the flat sky approximation.

While the flat sky approximation has been derived previously~\cite{Datta:2006vh,White:2017ufc,Castorina:2018nlb,Jalilvand:2019brk}, its accuracy has not been analysed in any detail\footnote{More precisely, in Ref.~\cite{Datta:2006vh} the authors claim that they find an accuracy of better than 1\% for redshift differences of $\De \nu/\nu_0= \delta \simeq 10^{-3}\ll \delta_0(z)\simeq 0.06$ at redshift $z=10$. We roughly agree with this as we shall see later in the present paper.}.
Doing this is the goal of the present paper.  
Our target accuracy for the total power spectrum is $\leq1\%$ for correlations at equal redshift, and $\leq 5\%$ for unequal redshifts. This accuracy should be maintained across the redshift interval $1\leq z\leq5$, and down to $10\lesssim \ell_{\min} \lesssim 100$. 

Let us compare this with the accuracy which can be achieved in upcoming observational surveys such as Euclid. At best, their error is dominated by the irreducible contributions from 
cosmic variance and shot noise leading to
\bea
\De C_\ell(z_1,z_2) = \frac{2}{f_{\rm sky}(2\ell +1)}\sqrt{C_\ell(z_1,z_1)C_\ell(z_2,z_2)}  +\frac{1}{\sqrt{n(z_1)n(z_2)}} \,. 
\eea
Here $f_{\rm sky}$ is the sky coverage of the survey and $n(z)$ is the number of galaxies in the redshift bin considered. The first term is cosmic variance~\cite{RuthBook} and the second term is the shot noise. If the number of galaxies is very high such that shot noise can be neglected (e.g. for the photometric survey of Euclid), for equal redshifts, $z_1=z_2$ this error becomes very small, of the order of 1\% and less for $f_{\rm sky} \sim 0.3$ and $\ell\gtrsim 300$. However, for $z_1\neq z_2$, we have that $|C_\ell(z_1,z_2)| \ll \sqrt{C_\ell(z_1,z_1)C_\ell(z_2,z_2)}$, depending on the redshift difference. Thus, the relative error never becomes very small. For significant redshift differences it is always at least 5\% or more. Therefore, an accuracy of 5\% is  considered good enough for unequal redshifts\footnote{Of course when $z_1\ra z_2$, the unequal $z$ spectra converge to the equal $z$ ones. To have a factor $|C_\ell(z_1,z_2)|/\sqrt{C_\ell(z_1,z_1)C_\ell(z_2,z_2)} < 0.2$, one typically has to require $|z_1-z_2|\gtrsim 0.5$. }.

In recent years, fully relativistic expressions for the fluctuations of the observed galaxy number counts and their spectra have been derived~\cite{2009PhRvD..80h3514Y,Yoo:2010ni,Bonvin:2011bg,Challinor:2011bk}. A detailed study of these spectra has shown~\cite{Bonvin:2011bg,Challinor:2011bk,Jeong:2011as,Yoo:2013tc,DiDio:2013bqa,DiDio:2013sea,Bonvin:2014owa,Camera:2014bwa,Montanari:2015rga,Tansella:2017rpi} that, in nearly all situations, the large scale relativistic terms are very small and can be neglected for percent level accuracy. Exceptions to this are very low redshifts, $z<0.1$, see~\cite{Tansella:2017rpi}, and very large angular scales, $\ell<10$. The latter are not very relevant, at least for single tracer analyses, due to cosmic variance and their importance is discussed in Refs.~\cite{Bruni:2011ta,Alonso:2015uua,Raccanelli:2015vla,Lorenz:2017iez}.
A new code significantly speeding up the cumbersome line of sight integrations has recently been developed~\cite{Schoneberg:2018fis}.

The remaining terms which are relevant on sub-horizon scales are the density, redshift-space distortion (RSD) and the lensing term. These are the terms which we investigate here and for which we derive simple approximations that can be computed rapidly, but are nevertheless accurate at the 0.5\% level or better for equal redshift correlations. For unequal redshifts our approximations for density and RSD are much less precise, but the lensing terms, which can be computed with the Limber approximation, dominate for large redshift differences.  In the past, the density and RSD terms have been computed mainly in Fourier space~\cite{Kaiser1987}. This is sufficient for small surveys in one redshift bin. 
We shall see that the flat sky approximation for density and RSD, while requiring a similar numerical effort, is not equivalent and is valid also at very large angular scales down to $\frac{\pi}{\theta}\simeq \ell=2$. 

The lensing term is an integral along the line of sight and cannot, without approximations, be represented in Fourier space (see~\cite{Tansella:2017rpi} for an attempt). As the truly measured quantities are directions and redshifts, it is most consistent and model independent to represent
the number count fluctuations as a function of direction and redshift. This is what we do in this work. When assuming a background cosmology, the redshift space correlation function can also be 
computed and may be more useful for the analysis of spectroscopic surveys~\cite{Tansella:2017rpi,Tansella:2018sld} within one redshift bin. However, for the very promising analysis of number counts from large photometric surveys, angular--redshift power spectra will most probably become the method of choice, since they are truly model independent.  Angular and redshift fluctuations are also simple to combine with shape measurements in order to derive galaxy-galaxy lensing cross-correlation spectra, see e.g.~\cite{Abbott:2017wau,Ghosh:2018nsm}.

The remainder of this paper is structured as follows: In the next section we introduce the flat sky approximation for density, RSD and lensing and we compare flat sky results with {\sc CLASS} results. We first present results for equal redshift correlations which are accurate to below $\sim0.5\%$ for $\ell\gtrsim10$ and then study unequal redshifts which are more problematic. In Section~\ref{s:Limber} we compare the flat sky and the Limber approximations and in Section~\ref{s:con} we summarize our findings and conclude. 

\subsection{Notation and Conventions}
We set the speed of light $c=1$ throughout. We consider a Friedmann universe with scalar perturbations only in longitudinal (Newtonian, Poisson) gauge,
\be
ds^2 = a^2(t)\left[-(1+2\Psi)dt^2 + (1-2\Phi)\de_{ij}dx^idx^j\right] \,.
\ee
We denote conformal time by $t$ and a derivative with respect to $t$ by an overdot. The conformal Hubble parameter is denoted by $\HH=\dot a/a$ while the physical Hubble parameter is $H=\HH/a$. 

\section{The Flat Sky Approximation}
\subsection{Generalities}
Linear perturbation theory gives the following expression for the number count fluctuations in direction $\bn$ at redshift $z$~\cite{Bonvin:2011bg,Challinor:2011bk,DiDio:2013bqa},
\be\label{e:De}
\De(\bn,z) = b(z)D(r(z)\bn,t(z)) +\frac{1}{\HH(z)}\dd_rV_r(r(z)\bn,t(z)) - 2(1-\ga(z))\ka(\bn,z) \,,
\ee
where RSD and lensing are included, but large scale relativistic effects are neglected. Here $b(z)$ is the linear galaxy bias which depends on the class of galaxies considered in the survey, 
$D$ is the density fluctuation (in comoving gauge), $V_r$ is the radial component of the velocity field (in longitudinal gauge) and $\ka$ is the convergence,
\be
2\ka(\bn,z) = \De_{\Stwo}\int_0^{r(z)}\frac{dr'(r(z)-r')}{r(z)r'}(\Psi(r'\bn,t_0-r')+\Phi(r'\bn,t_0-r'))\,,
\ee
where $ \De_{\Stwo}$ denotes the Lapace operator on the 2-sphere, i.e. with respect to $\bn$. 
The function $\ga(z)$ is the luminosity bias which is given by the logarithmic derivative of the observed galaxy population at the flux limit of the given survey,
\be
\ga(z,F_{\rm lim}) \equiv -\left.\frac{\partial\log N(z,F> F_{\rm lim})}{\partial\log F}\right|_{F_{\rm lim}} \,.
\label{e:s_mlim}
\ee
Here $N$ denotes  the mean density of galaxies which are seen with a flux $F>F_{\lim}$ from redshift $z$, i.e. the  density of  galaxies with luminosity $ L> L_{\rm lim}$. The  luminosity is related to the flux as usual via the luminosity distance $D_L$, $F= L/(4\pi D_L(z)^2)$. 
Clearly, this function is  survey-dependent, but $\ga(z)$ is also directly observable and does not depend on the background cosmology (which determines e.g. $D_L(z)$).

The number count fluctuation can be expanded in spherical harmonics,
\bea\label{e:sphf}
\De(\bn,z) &=& \sum_{\ell m}a_{\ell m}(z)Y_{\ell m}(\bn)\\
a_{\ell m}(z) &=& \int_{\Stwo} \De(\bn,z)Y^*_{\ell m}(\bn)d\Om_{\bn} \,, \label{e:sphb}
\eea
and the angular redshift power spectrum is given by
\be
\langle a_{\ell m}(z)a^*_{\ell' m'}(z')\rangle  =C_{\ell}(z,z')\de_{\ell\ell'}\de_{mm'} \,.
\ee
Like for the CMB, the Kronecker-deltas are a consequence of statistical isotropy.

In the flat sky approximation we replace the direction $\bn$ by $\bn = \boe_z +\bal$ where $\bal$ lives in the plane normal to $\boe_z$, the flat sky. The direction $\boe_z$ is a reference direction out to the center of our survey. In the flat sky, $\bell$ is the dimensionless variable conjugate to $\bal$ and the spherical harmonic transform define in Eq.~(\ref{e:sphb}) of an arbitrary variable $X$  becomes a 2D Fourier transform,
\bea
a^X(\bell,z) \simeq \frac{1}{2\pi}\int d^2\al e^{i\bell\cd\bal}X(\bal,z)\,, \qquad  &&
X(\bal,z) \simeq \frac{1}{2\pi}\int d^2\ell e^{-i\bell\cd\bal}a^X(\bell,z) \,.\label{e:Xbal1}
\eea

Let us first consider a variable $X(\bx,t)$ defined in all of space at any time with transfer function $T_X(k,z)$ so that $X$ is given by
\bea
X(\bk,z) &=&T_X(k,z)\RR(\bk)\,,\\
X(\bx,z) &=&\frac{1}{(2\pi)^3}\int d^3ke^{-i\bx\cd\bk}T_X(k,z)\RR(\bk) \,. \label{e:Xofx}
\eea
Here $\RR(\bk)$ is the initial curvature fluctuation after inflation. Its power spectrum is defined by
\bea \label{e:Pdim}
\langle \RR(\bk)\RR^*(\bk')\rangle &=& (2\pi)^3\de(\bk-\bk')P_{\RR}(k) \,,\\
\frac{k^3}{2\pi^2}P_{\RR}(k) &=&  \PP_{\RR}(k)\,.   \label{e:Padim}
\eea
The normalization of the dimensionless power spectrum $\PP_{\RR}$ is such that the correlation function of $\RR$ in real space is simply
\be
\langle\RR(\bx)\RR(\by)\rangle = \int_0^\infty\frac{dk}{k}j_0(kr) \PP_{\RR}(k)\,,  \qquad r=|\bx-\by|\,,
\ee
without any pre-factor. Here $j_0$ is the spherical Bessel function of order $0$, see~\cite{Abram}. The scalar perturbation amplitude $A_s$ and  the scalar spectral index $n_s$ are defined by
\be
\PP_\RR(k) = A_s(k/k_*)^{n_s-1} \,,
\ee
where $k_*$ is called the pivot scale.
For numerical examples in this paper we shall use the Planck values\cite{Aghanim:2018eyx} for  $k_*=0.05/$Mpc,\\

\bea
\log(10^{10}A_s) &=& 3.043 \nonumber\\
\qquad n_s&=&0.9652 \,. \nonumber
\eea
\begin{flushleft}
Inserting $X(\bal,z)=X(r(z)(\boe_z+\bal),z)$
in Eq.~\eqref{e:Xofx}  we find
\end{flushleft}
\be\label{e:Xbal}
X(\bal,z)= \frac{1}{(2\pi)^3}\int d^3ke^{-ir(z)(\boe_z+\bal)\cd\bk}T_X(k,z)\RR(\bk) \,.
\ee
Comparing this with Eq.~\eqref{e:Xbal1} and identifying $\bk = k_\pa\boe_z+\bell/r(z)=k_\pa\boe_z+\bk_\perp$, we find
\bea \label{e:aXlz}
a^X(\bell,z) &=& \frac{1}{(2\pi)^2}\int d^3ke^{-ir(z)(\boe_z+\bal)\cd\bk}T_X(k,z)\RR(\bk)\de(\bell-r(z)\bk_\perp) \\
&=& \frac{1}{(2\pi)^2r(z)^2}\int_{-\infty}^\infty dk_\pa e^{-ir(z)k_\pa}T_X(k,z)\RR(\bk)
\eea
where $\bk = k_\pa\boe_z+\bell/r(z)$ and $k =\sqrt{k_\pa^2 +\ell^2/r(z)^2}$.

\subsection{Equal redshift correlations}
Let us first correlate two variables $X$ and $Y$ at the same redshift, $a^X(\bell,z) $ and $a^{Y\,*}(\bell',z) $. Using Eq.~\eqref{e:aXlz} we obtain
\bea
\langle a^X(\bell,z)a^{Y*}(\bell',z)\rangle  &=&\frac{1}{2\pi}\int d^3kP_{\cal R}(k)\de^2(\bell-r(z)\bk_\perp)\de^2(\bell'-r(z)\bk_\perp)T_X(k,z)T^*_Y(k,z) \hspace*{1cm}\\
 &=& \de^2(\bell-\bell')\frac{1}{\pi r(z)^2}\int_0^\infty dk_\pa P_{\cal R}(k)T_X(k,z)T^*_Y(k,z) \,,\\
C^{XY}_\ell(z,z) &=& \frac{1}{\pi r(z)^2}\int_0^\infty dk_\pa P_{\cal R}(k)T_X(k,z)T^*_Y(k,z)\,. \label{e:Clflat1}
\eea
The situation is  more complicated when we consider different redshifts $z\neq z'$, and we defer a discussion of this case to  Section~\ref{s:uneq}.

To determine the number counts we have to apply our formalism to the density fluctuation in comoving gauge, $D$, the the radial component of the velocity, $V_r$ and to the Weyl potential, $\Psi_W=(\Psi+\Phi)/2$. The latter then has to be integrated over the lightcone in order to obtain $\ka$:
\bea \label{e:akaelde}
 a^D(\bell,z) &=& \int \frac{d^3k}{(2\pi)^2}\de(\bell-r(z)\bk_\perp)e^{-ik_\pa r(z)}T_D(k,z){\cal R}(\bk)\\
a^\text{rsd}(\bell,z) &=&\HH^{-1}\int \frac{d^3k}{(2\pi)^2}\de(\bell-r(z)\bk_\perp)e^{-ik_\pa r(z)}\frac{k^2_\pa}{k} T_V(k,z){\cal R}(\bk) \,,  \label{e:akaelV}\\
\label{e:akael}
a^\ka(\bell,z) &=& 2(1-\ga(z))\ell^2\int \frac{d^3k}{(2\pi)^2}\int_0^{r(z)}dr\frac{r(z)-r}{r(z)r}\de(\bell-r(z)\bk_\perp )e^{-ik_\pa r}T_{\Psi_W}(k,z(r)){\cal R}(\bk) \,. \hspace*{1cm}\label{e:alzka}
\eea
For Eq.~\eqref{e:akaelV} we have used $\bV(\bk,z) =i\hat\bk T_V(k,z)\RR(\bk)$, hence $(\dd_rV_r)(k)= (k^2_\pa/k)T_V(k,z)\RR(\bk)$. In Eq.~\eqref{e:akael}, $z(r)$ is the redshift of the comoving distance $r$, i.e. $r(z(r))\equiv r$.

We now consider each term, first by itself and then its correlation with the other contributions. To compute a spectrum numerically we use the Planck values of $k_*$, $A_s$ and $n_s$ for $P_\RR(k)$ and employ the numerical transfer function from {\sc class} for the variable $X$. The remaining integral Eq.~\eqref{e:Clflat1}  is then computed with a simple Python code. For the lensing term this would lead to triple 
or double integrals and we therefore perform additional simplifications, as set out in Section~\ref{s:len}.

\subsubsection{Density Fluctuations}\label{s:den}

\begin{figure}[h!]
\includegraphics[width=0.95\linewidth]{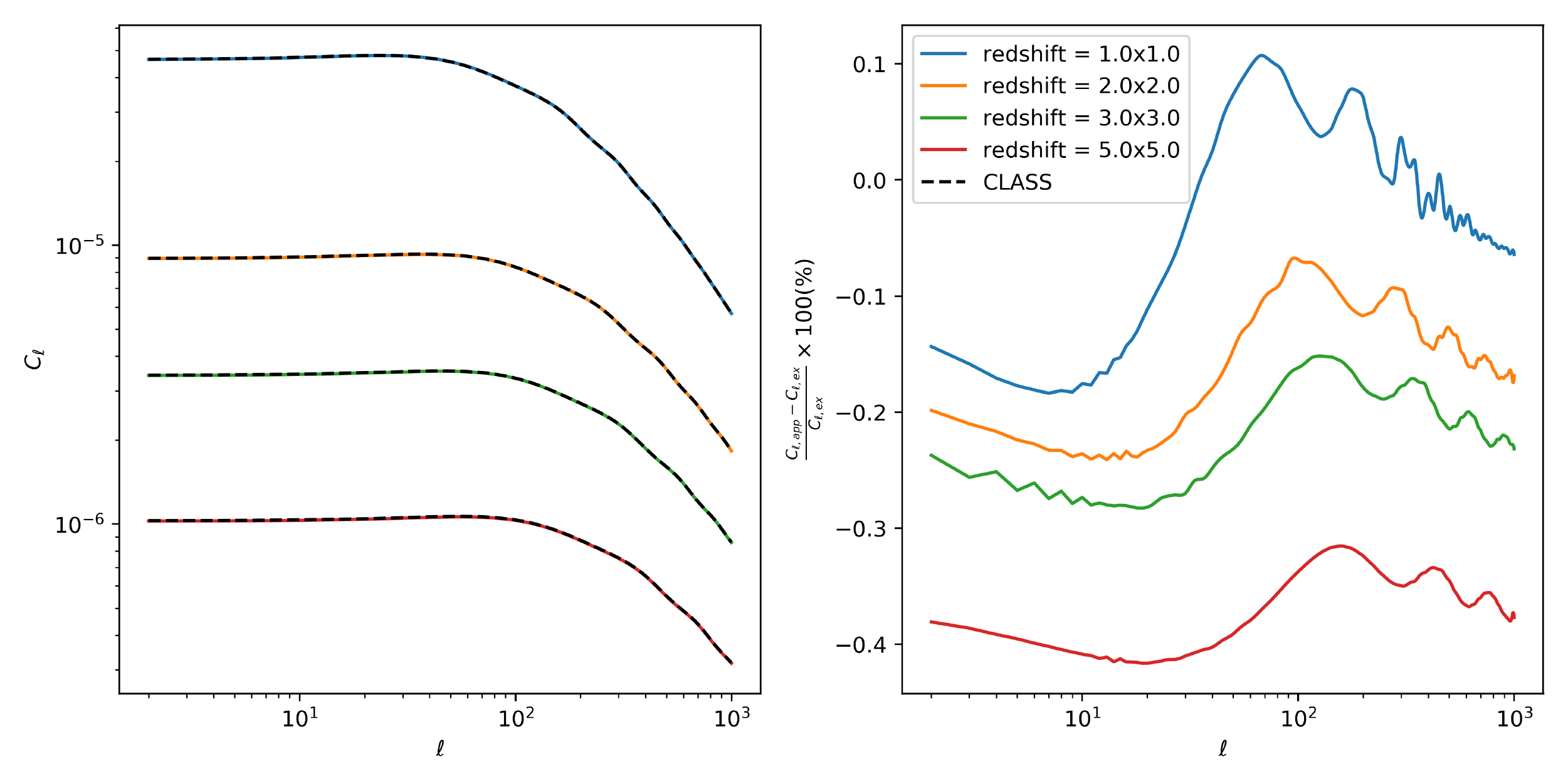}
\caption{\label{f:density} Left: the density in the flat sky approximation (solid) compared to the {\sc class} result (dashed) at redshifts   $z=1,~2,~3$ and $5$ from top to bottom. Right: the relative differences in percent.} 
\end{figure}

In Fig.~\ref{f:density} we plot the density power spectrum at redshifts $z=1,~2,~3$ and $5$. The solid lines present our approximation Eq.~\eqref{e:Clflat1} with $X=Y=D$ while the dashed lines are the numerical result from {\sc class}. The agreement is clearly excellent. To be  more quantitative we also show the relative differences  in the right panel. The difference is  about 0.4\% for $z=5$,  about 0.1\% for $z=1$ and typically 0.2\% for $1<z\leq 3$. Note that  0.1\% is roughly the accuracy of the {\sc class} code itself, hence our agreement is as good as we can expect.

\subsubsection{RSD}\label{s:rsd}
Here we repeat the same analysis for RSD, where we have
\be
X(z,\bk)=k_\pa V_r/\HH= \frac{k}{\HH(z)}\left(\frac{k_\pa}{k}\right)^2T_V(k,z)\RR(\bk)\,.
\ee

\begin{figure}[!ht]
\includegraphics[width=0.95\linewidth]{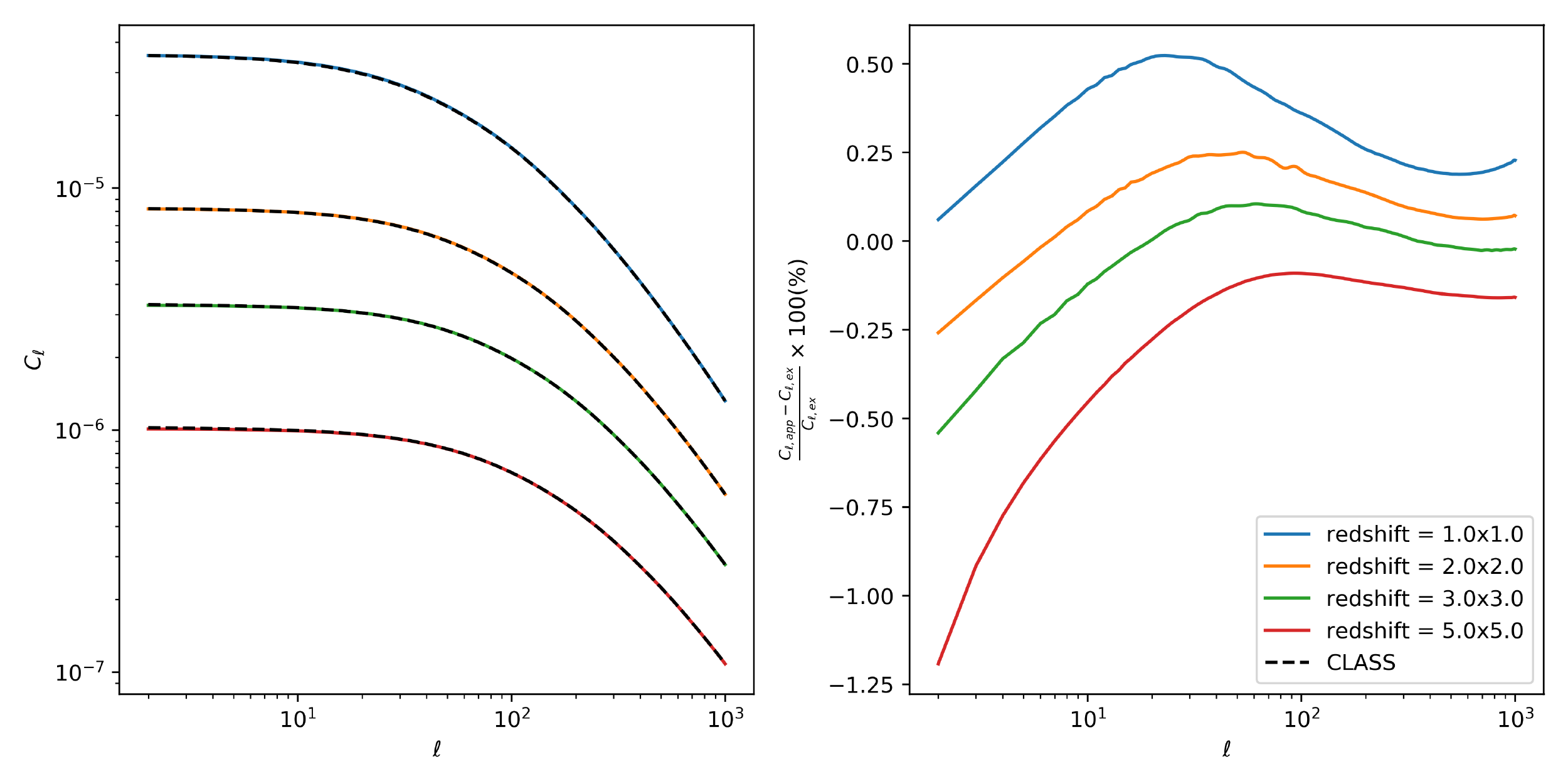}
\caption{\label{f:rsd} Left: the RSD term in the flat sky approximation (solid) compared to the {\sc class} result (dashed) at redshifts   $z=1,~2,~3$ and $5$ from top to bottom. Right: the relative differences in percent.} 
\end{figure}

In Fig.~\ref{f:rsd} we plot the RSD power spectrum at redshifts $z=1,~2,~3$ and $5$. The solid lines present our approximation Eq.~\eqref{e:Clflat1} with $X=Y=$ RSD while the dashed line are the numerical result from {\sc class}. 
Again we see very good agreement with the numerical result from {\sc class}. At low $\ell$ the approximation degrades somewhat, but the error is never larger than about 1\%. In general the difference is 0.5\% or less.

Again at high redshift our result is slightly below the  {\sc class} value while at low redshift it is slightly above.

\subsubsection{Lensing}\label{s:len}

To obtain the lensing term we write the flat sky approximation as
\bea
\langle  a^\ka(\bell,z) a^{\ka*}(\bell',z')\rangle &=&\de^2(\bell-\bell')\frac{2(1-\ga(z))(1-\ga(z'))\ell^4}{\pi}\int_{-\infty}^\infty \hspace{-0.23cm}dk_\pa \int_0^{r(z)}\hspace{-0.2cm}dr\int_0^{r(z')}\hspace{-0.2cm}dr'  \nonumber \\ &&
\hspace*{-0.3cm}\times  \frac{(r(z)-r)(r(z')-r')}{r(z)r(z')(rr')^2}e^{ik_\pa(r-r')}P_{\cal R}(k)
 T_{\Psi_W}(k,z(r))T^*_{\Psi_W}(k,z(r')) \,.~  \label{e:ka-full}
\eea
 Eq.~\eqref{e:ka-full}
 is a triple integral of a rapidly oscillating function and hence very time-consuming numerically.
To simplify it, we  integrate Eq.~\eqref{e:ka-full} over $k_\pa$ neglecting the dependence of the transfer functions and the power spectrum on $k_\pa$, i.e. simply  setting $k_\pa=0$ in the expression  $k=\sqrt{k_\pa^2+(\bell/r)^2}$. The integral of the exponential over $k_\pa$ then yields a $\de$-function  in the resulting expression. This corresponds to setting
$$\int dk_\pa f(r',r,k)\exp(ik_\pa (r-r')) \simeq 2\pi  f(r,r,|\bk_\perp|)\de(r-r')
 $$ 
 which is a good approximation for a slowly varying function $f(r,r',k)$ and for $\ell/r\gg k_\pa$. The integral over $r'$  then simply eliminates the $\de$-function and we obtain
 \be\label{e:Cl-ka}
 C^\ka_\ell(z,z') = 4\ell^4(1-\ga(z))(1-\ga(z'))\int_0^{r_{\min}}\!\!\!\!\!dr\frac{(r(z)\!-\!r)(r(z')\!-\!r)}{r(z)r(z')r^{4}}P_{\cal R}(k)|T_{\Psi_W}(k,z(r))|^2 \,,
 \ee
where now $k=\ell/r$ and $r_{\min}=\min\{r(z),r(z')\}$.
As we shall see in Section~\ref{s:Limber}, Eq.~\eqref{e:Cl-ka} simply corresponds to the Limber approximation~\cite{Limber:1954zz,LoVerde:2008re} which is often used for lensing. Eq.~\eqref{e:Cl-ka} is a single integral of a positive definite slowly-varying function which does not pose any problem and can be calculated with a simple Python code.

\begin{figure}[ht]
\includegraphics[width=0.95\linewidth]{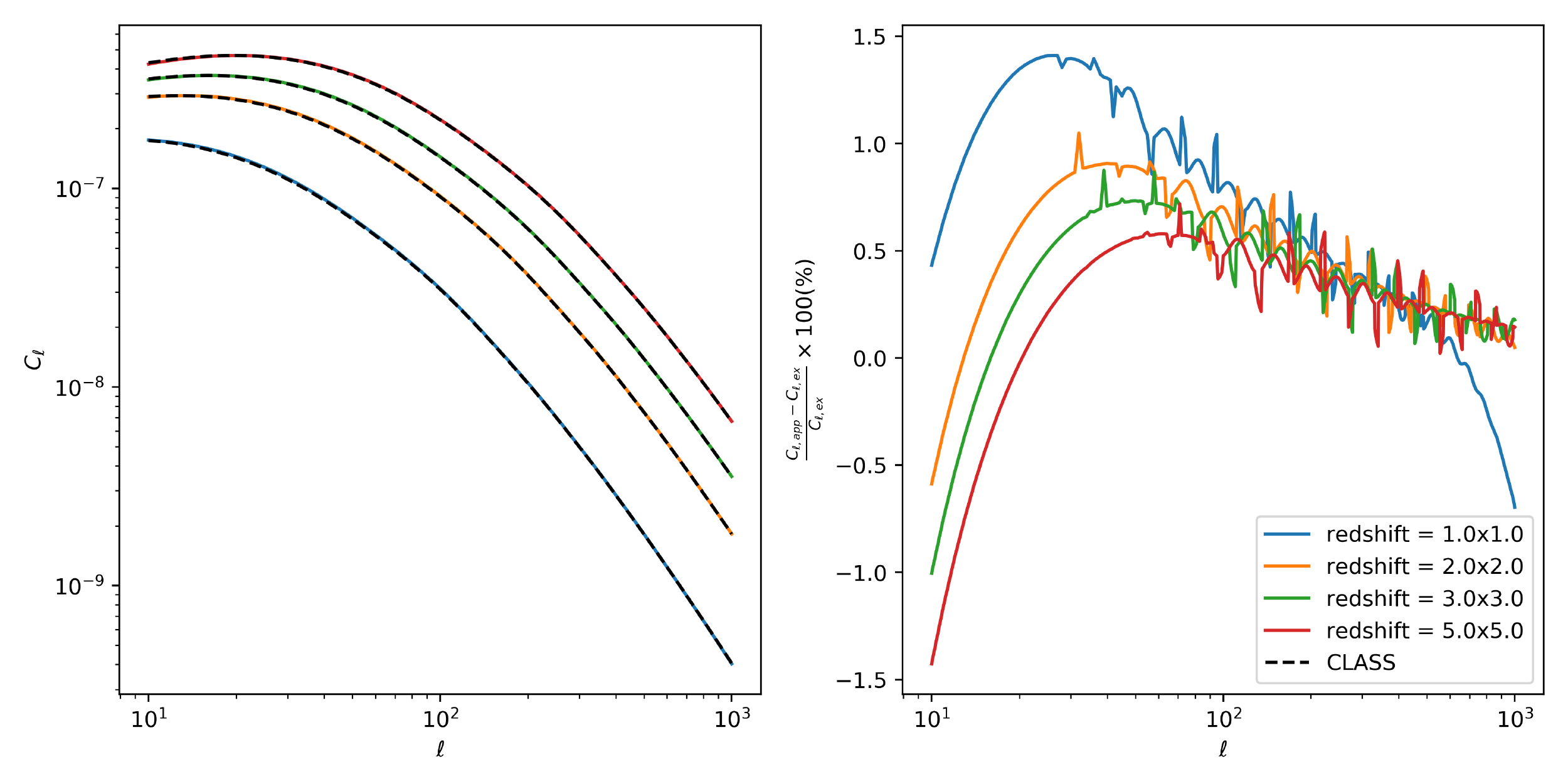}
\caption{\label{f:ka} Left: the lensing term in the flat sky approximation (solid) compared to the {\sc class} result (dashed) at redshifts   $z=1,~2,~3$ and $5$ from  bottom to top. Right: the relative differences. We have set $\ga(z)=0$ in this numerical evaluation.} 
\end{figure}

In Fig.~\ref{f:ka} we show $C^\ka_\ell(z,z)$ for the redshifts $z=1,~2,~3,~5$.  Our approximation Eq.~\eqref{e:Cl-ka}  is excellent for $\ell> 20$ where the relative difference are typically about 0.5\% and only for $z=1$ larger than  1\%.
The same is true for unequal redshifts which are shown in Fig.~\ref{f:ka-x}. The agreement is bad at low $\ell\leq 10$, but this is not so surprising as for small $\ell$, neglecting $k_\pa$ with respect to $k_\perp = \ell/r$ is certainly not a good approximation. For $z=1$ the error also rises above 1\% for $20<\ell<50$. But  lensing from $z=1$ is very subdominant (compare the amplitudes in Figs.~\ref{f:density}  and 
\ref{f:ka}) so that this error does not contribute significantly to the total error budget.

Even though this approximation is much better than the flat sky approximation for unequal redshifts which we discuss below, it cannot capture the behavior of the lensing term at low $\ell$. Nevertheless, we shall see that the low $\ell$ contribution from the lensing terms is subdominant so that we can still achieve a good approximation for the power spectrum of the full number count $C_\ell^\De(z,z)$. Note that while density and RSD decrease with increasing redshift, the situation is reversed for the integrated lensing term.  While at $z=1$, the lensing is about 100 times smaller than the density term, at redshift $z=5$ it is only about two times smaller at low $\ell$.

\subsubsection{Cross-Correlations}
The number count expression Eq.~\eqref{e:De} implies that the full number count power spectrum is given by
\bea
C_\ell^\De(z,z') &=& b(z)b(z')C_\ell^D(z,z') + b(z)C_\ell^{D,\text{rsd}} (z,z')+ b(z')C_\ell^{\text{rsd},D}(z,z') +C_\ell^{\text{rsd}}(z,z')  \nonumber\\ && 
+b(z)(1-\ga(z'))C_\ell^{D,\ka}(z,z') +(1-\ga(z))b(z')C_\ell^{\ka, D}(z,z') +(1-\ga(z'))C_\ell^{\text{rsd},\ka}(z,z')   \nonumber\\ &&  +(1-\ga(z))C_\ell^{\ka, \text{rsd}}(z,z')+(1-\ga(z))(1-\ga(z'))C_\ell^{\ka}(z,z')
\label{e:Cl-tot}
\eea
For simplicity and in order to be as model independent as possible, we set $b(z)=1$ and $\ga(z)=0$ in the following, but they are easily re-introduced for any specific example.
In this section we show numerical examples of the correlation spectra, $C_\ell^{X,Y}(z,z)$ for $X\neq Y$. As a consequence of the continuity equation, the velocity transfer function is very simply related to the density transfer function via
\be\label{e:growth}
T_V(k,z)= f(z)T_D(k,z)\,, \qquad  f(z) = -\frac{d\log D_1(z)}{d\log(1+ z)} \,,
\ee
where $D_1(z)$ is the growth function of density perturbations. In a matter-dominated universe $D_1\propto1/(1+z)$, while during dark energy domination $D_1$ grows slower and tends to a constant as $\Om_m\ra 0$.
With this the correlation $C_\ell^{D,\text{rsd}} (z,z)$ simply becomes
\be
C^{D,\text{rsd}}_\ell(z,z) \simeq \frac{f(z)}{\pi r(z)^2\HH(z')}\int_0^\infty dk_\pa \frac{k_\pa^2}{k}P_{\cal R}(k)T_D(k,z)T_D^*(k,z)\,.
\ee

\begin{figure}[ht]
\includegraphics[width=0.95\linewidth]{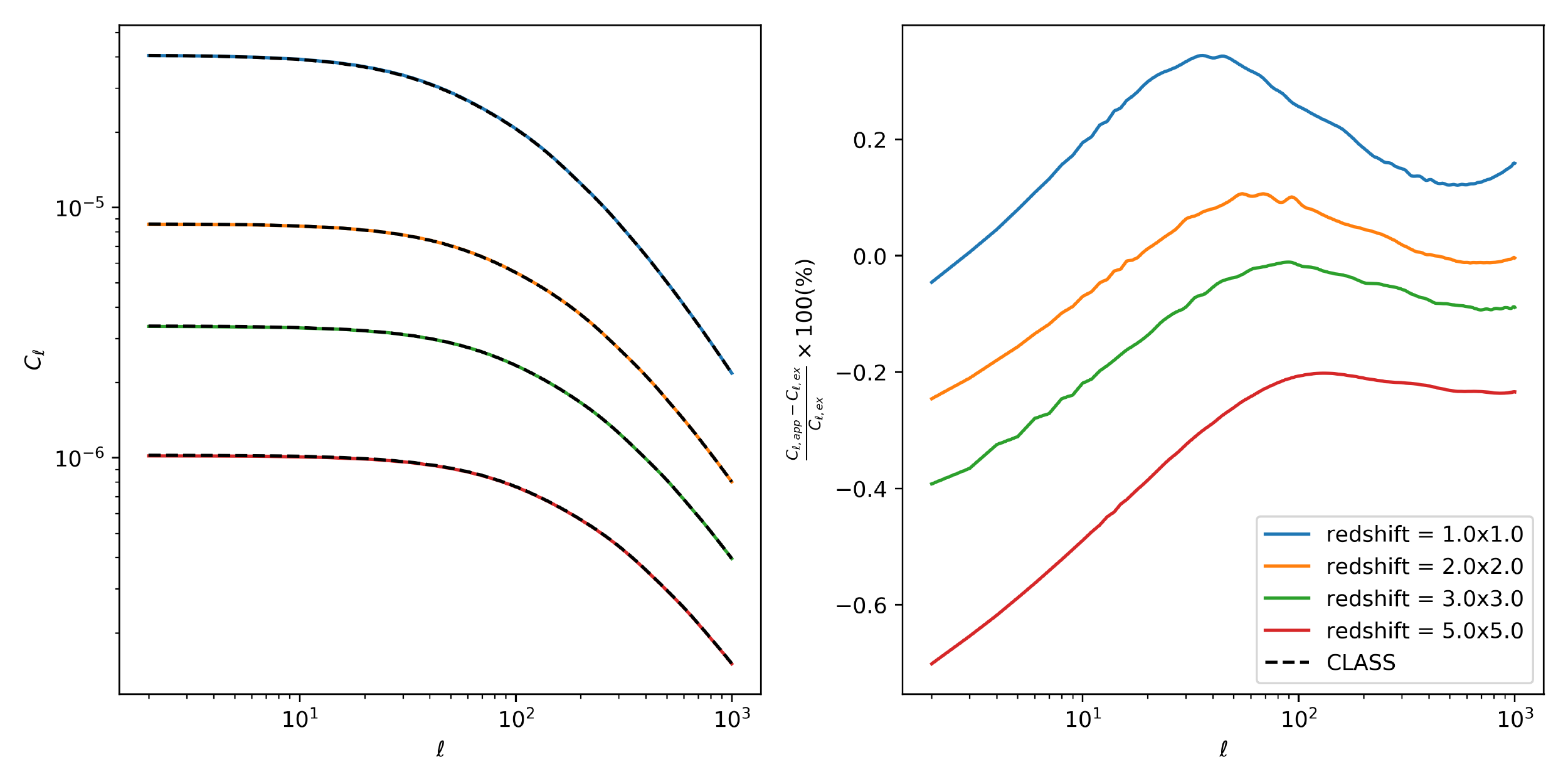}
\caption{\label{f:DRSD} Left: the density-RSD cross spectrum in the flat sky approximation (solid) compared to the {\sc class} result (dashed) at redshifts   $z=1,~2,~3$ and $5$ from top to bottom. Right: the relative differences.} 
\end{figure}
In Figure~\ref{f:DRSD} we show some examples of the density-RSD cross-correlation spectrum for equal redshifts.
Not surprisingly, the errors are like the ones for density or RSD terms, i.e. for $\ell\geq 10$ never larger than 0.5\% and largest for high redshift and low $\ell$. More precisely, the largest error at $\ell=2$ and $z=5$ is 0.7\%.

Let us consider the lensing-density cross-correlation next. Inserting Eqs.~(\ref{e:akaelde},\ref{e:akael}) in Eq.~\eqref{e:Clflat1} we obtain
\be
C^{D,\ka}_\ell(z,z') \simeq \frac{-\ell^2}{\pi }\int_{-\infty}^\infty dk_\pa\int_0^{r(z')}\!~\!\! dr \frac{r(z')-r}{r(z')rR^2} P_{\cal R}(k)T_D(k,z)T_{\Psi_W}^*(k,z(r))\exp(ik_\pa (r(z)-r))\,.
\ee 
Here $R=\sqrt{r(z)r}$, hence we cannot take $ P_{\cal R}(k)$ in front of the $r$-integration, as $k=\sqrt{k_\pa^2+\ell^2/R^2}$. We perform the same simplification as in the lensing integral. We neglect the dependence on $k_\pa$ in  $k$ and integrate the exponential over $k_\pa$ which then yields $2\pi\de(r(z)-r)$ so that we end up with
\be
\label{e:Cl-kd}
C^{D,\ka}_\ell(z,z') \simeq \left\{\begin{array}{cc}
-2\ell^2\frac{r(z')-r(z)}{r(z')r(z)^3} P_{\cal R}(\ell/r(z))T_D(\ell/r(z),z)T_{\Psi_W}^*(\ell/r(z),z)
 & \mbox{ if } z<z' \\
 0  & \mbox{ if } z\geq z' \,.\end{array}
\right.
\ee

Thus, for equal redshifts in this approximation, the contribution from the lensing-density cross-correlation vanishes. This is well-justified as the spectrum for this term is between two and three orders of magnitude smaller than the density or RSD autocorrelations individually, and can safely be neglected for equal redshifts.

\begin{figure}[ht]
\includegraphics[width=0.95\linewidth]{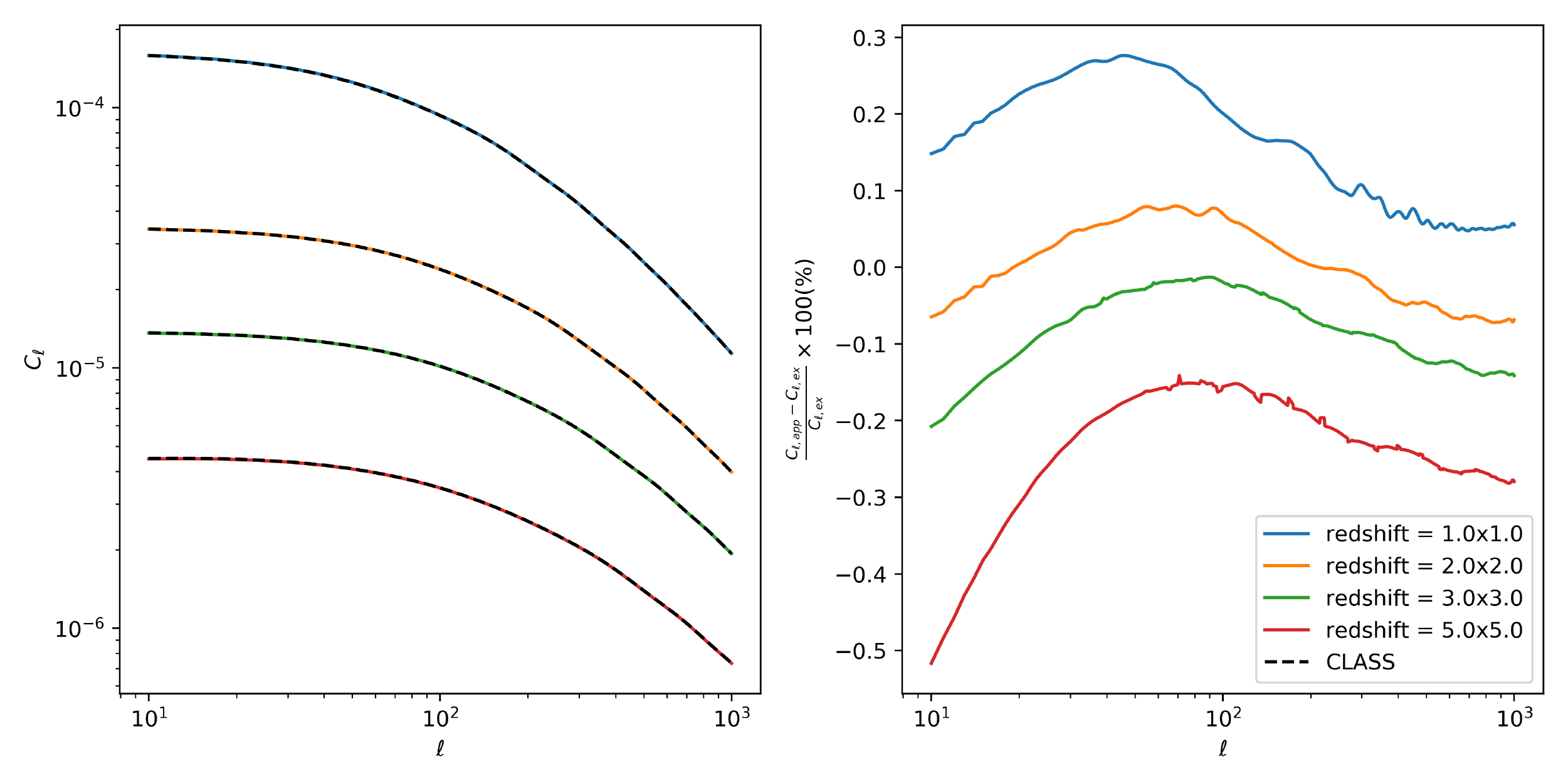}
\caption{\label{f:tot} Left: the full power spectrum, $C_\ell^\De(z,z)$ in the flat sky approximation (solid) compared to the {\sc class} result (dashed) at redshifts   $z=1,~2,~3$ and $5$  from top to bottom. Right: the relative differences.} 
\end{figure}

This same approximation, due to the factor $k_\pa$ in the transfer function, yields
\be
C^{\text{rsd},\ka}_\ell(z,z') \simeq 0 \,.
\ee
In Figure~\ref{f:tot} we show the total power spectrum result. The precision is always better than about 0.5\%.  Hence, neglecting the lensing-density and lensing-RSD terms does not degrade the accuracy. At the highest redshift, $z=5$, lensing contributes about 10\% to the total result, while at $z=1$ it drops to about 0.1\% and thus below the accuracy of our approximation.

\subsection{Unequal redshift correlations}\label{s:uneq}
\begin{figure}[ht]
\includegraphics[width=0.9\linewidth]{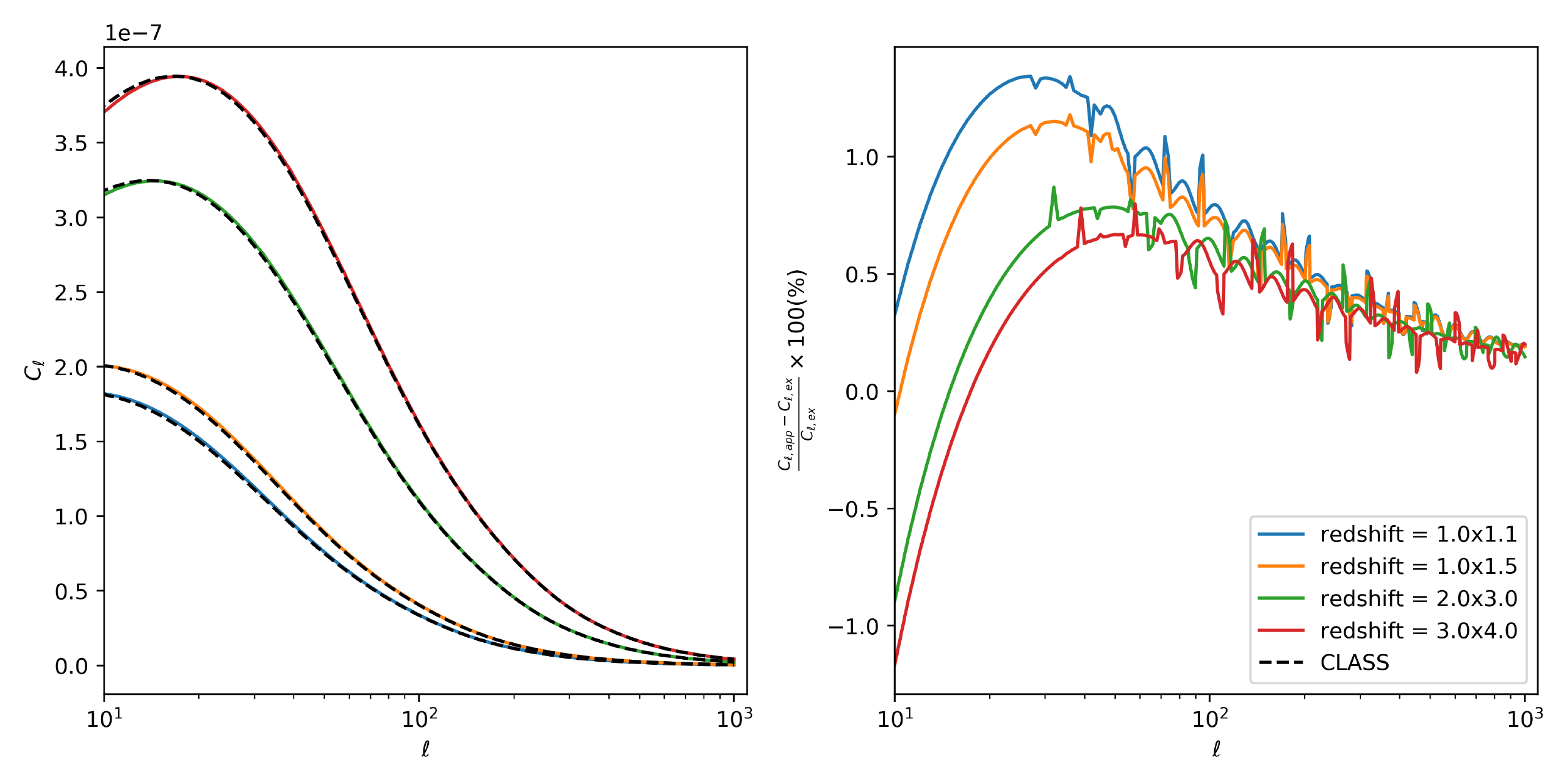}
\caption{\label{f:ka-x} Left: the lensing term in the flat sky approximation (solid) compared to the {\sc class} result (dashed) at redshifts   $(z,z')=(1,1.1)$, $(z,z')=(1,1.5)$, $(z,z')=(1,2)$, $(z,z')=(3,4)$ from bottom to top. Right: the relative differences.} 
\end{figure}

\begin{figure}[ht!]
\includegraphics[width=0.9\linewidth]{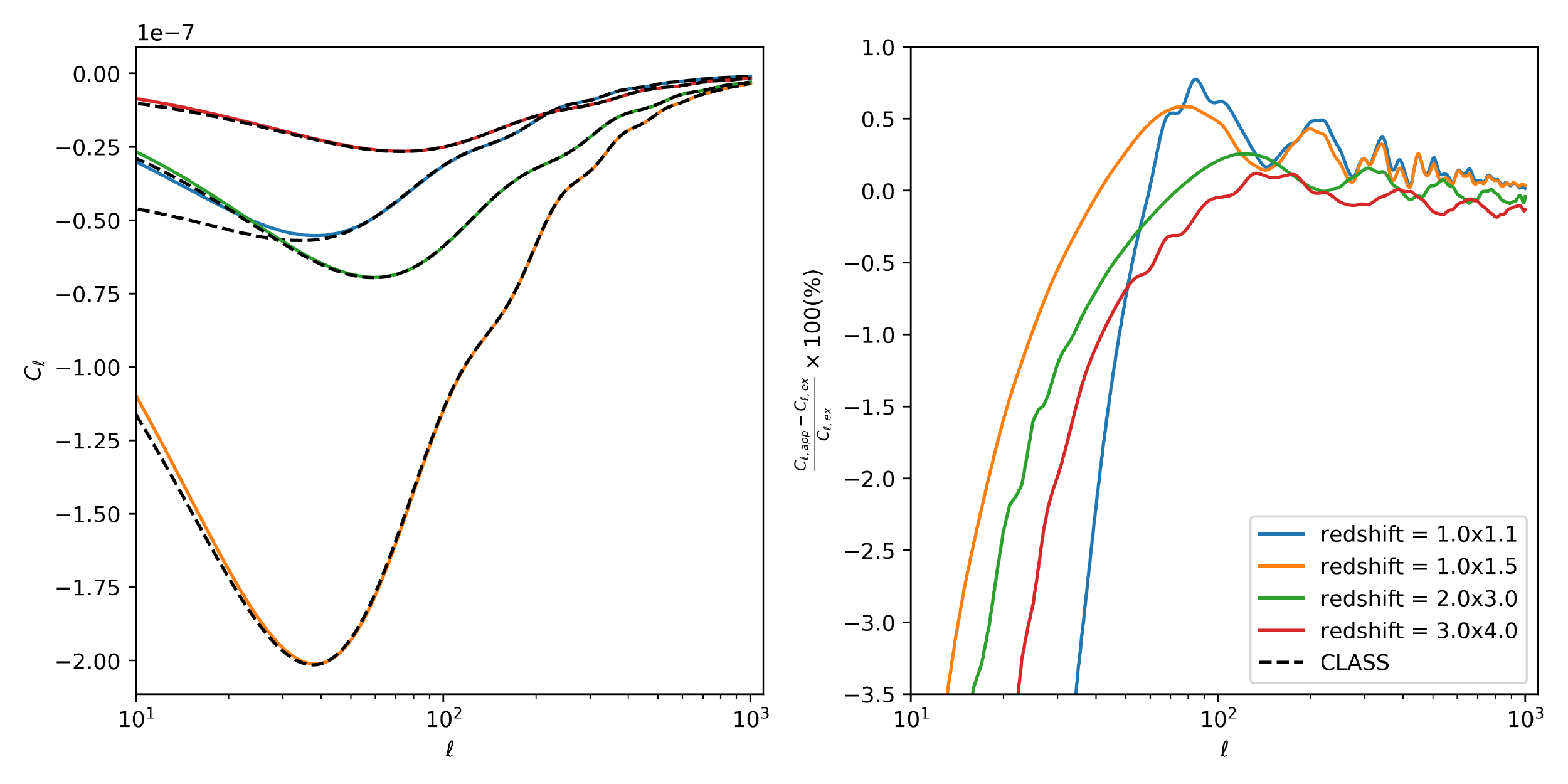}
\caption{\label{f:Dka-x} Left: the density-lensing cross spectrum in our approximation (solid) compared to the {\sc class} result (dashed) at redshifts   $(z,z')=(1,1.1),~(z,z')=(1,1.5),~(z,z')=(1,2),~(z,z')=(3,4)$ from top to bottom. Right: the relative differences.} 
\end{figure}

\begin{figure}[ht!]
\includegraphics[width=0.9\linewidth]{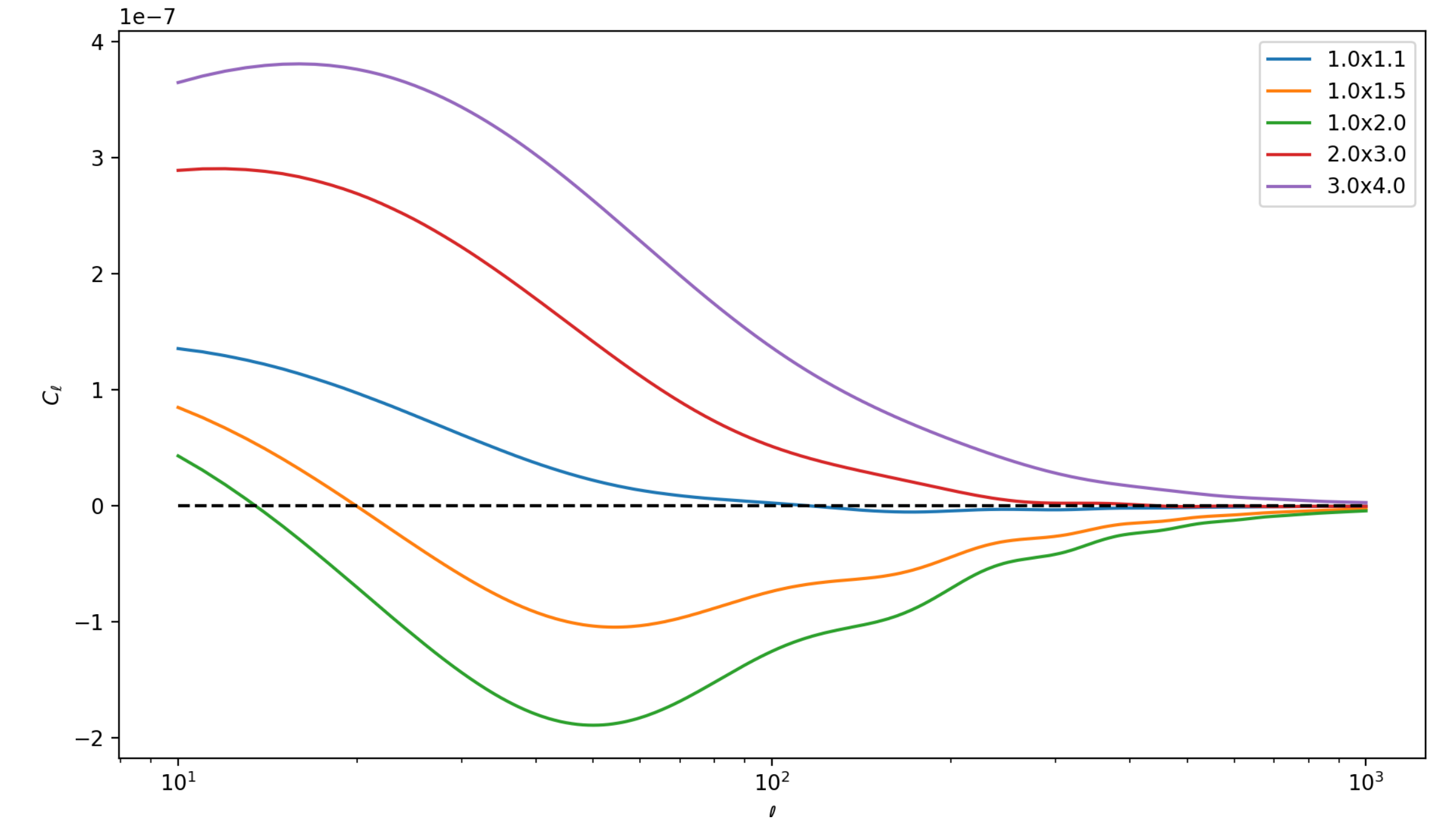}
\caption{\label{f:nonstd} The sum of the lensing-lensing and density-lensing cross spectra as produced by {\sc class} (solid) at redshifts   $(z,z')=(1,1.1),~(z,z')=(1,1.5),~(z,z')=(1,2),~(z,z')=(3,4)$, with a dashed line indicating zero signal.} 
\end{figure}

Let us also consider unequal redshifts.  For lensing and lensing-density cross-correlations we simply use approximations Eqs.~(\ref{e:Cl-ka},\ref{e:Cl-kd}).  The results are shown in Figs.~\ref{f:ka-x} and \ref{f:Dka-x}. While the lensing-lensing term is growing with redshift, the amplitude of the density-lensing term is more complex: the density decreases with increasing redshift while lensing increases. These two competing effects lead to a maximal signal (in amplitude, the sign of this term is always negative) at $(z,z')=(1,1.5)$. At $(z,z')=(3,4)$ the signal is smallest, while $(z,z')=(1,1.1)$ yields the second smallest signal (for $\ell>40$) and $(z,z')=(2,3)$ is the second largest signal. More precisely, the signals for $(z,z')=(1,1.5)$ and $(z,z')=(2,3)$ cross at $\ell\simeq 30$.  Note also that for $z\geq 2$, the positive lensing-lensing term always dominates over the negative density-lensing term, while for  $(z,z')=(1,1.5)$ the density-lensing term is larger for $\ell\gtrsim 100$ and for  $(z,z')=(1,1.1)$  it dominates already for $\ell\gtrsim 60$. This can be seen in Figure \ref{f:nonstd}, which shows the sum of these two terms.

The precision of the lensing-lensing approximation at different redshifts is as good as the 
one at equal redshifts, namely on the order of 1\% for $10\leq \ell\leq 100$ and around $0.5$\% for $\ell>100$.  For high redshifts, $z\geq 3$ this precision is reached already at lower $\ell$.

The density-lensing cross-correlations are  less accurate than the pure lensing term,  at low $\ell$. The reason here is that Limber approximation is less accurate for this case. This is especially true at low redshift, $z=1$ where for $\ell<60$ the error is larger than 1\% and it exceeds 4\% for $\ell<20$. For higher redshifts the accuracy is better and an accuracy of 0.5\% can be achieved for $\ell>50$. For $\ell>50$, the density-lensing cross-correlation is as accurate as the pure lensing term.
Comparing Figs~\ref{f:ka-x} and~\ref{f:Dka-x} , we see that the pure lensing term actually dominates in all the cases presented in these figures. Only at very low redshift, $z<1$, does the density-lensing term become larger. We have checked that the RSD-lensing term always remains very subdominant and we neglect it in our approximation.

Note also that these unequal-redshift correlations are always at least one or two orders of magnitude smaller than the full equal redshift result, which is dominated by the density and RSD terms. Therefore, the signal to noise of individual unequal-redshift terms is always very small. On the other hand, their number scales quadratically with the number of redshift bins and they can become relevant if we consider more than 10 bins. In Fig.~\ref{f:nonstd} we plot the sum of the two lensing terms.  Since the lensing-lensing term is positive while the lensing density term is negative, there can be significant cancellation. Especially, for $(z,z')=(1,1.1)$, the total signal nearly vanishes for $\ell>100$. For $(z,z')=(2,3)$ the same cancellation is effective for $\ell>200$. This second cancellation is especially relevant as at this redshift difference the lensing term usually dominates. Note, however that this cancellation depends on the magnification bias $\ga(z)$ since the density-lensing is proportional to $1-\ga(z)$ while the lensing-lensing term has the pre-factor $(1-\ga(z))(1-\ga(z'))$.

For the density and RSD, the so-called standard terms, unequal redshifts are  much less straightforward. The first difficulty is the following fact:
in real space, the correlation function for unequal redshifts is a function of $r\bal-r'\bal'$ where $r=r(z)$ and $r'=r(z')$. For $z\neq z'$ this  is not proportional to   $\bal-\bal'$. Therefore, upon Fourier transforming, we will not obtain a delta function $\de(\bell-\bell')$, since we break the flat sky analog of statistical isotropy. The reason for this is that in principle we now consider correlations of functions that live on two different skies: one at comoving distance $r(z)$ and the other at $r(z')$. In order to restore statistical isotropy we have to project them onto one sky at some fiducial common distance $R$.

 To do this we introduce the angles $\tilde\bal=\bal r/R$ and $\tilde\bal'=\bal' r'/R$, so that $r\bal-r'\bal' = R(\tilde\bal-\tilde\bal')$. Isotropy is now equivalent to translation invariance in the $\tilde\bal$ plane. Furthermore, we can write
\bea
a^X(\bell,z) &\simeq& \frac{1}{2\pi}\int d^2\tilde\al e^{-i\bell\cd\tilde\bal}X(\tilde\bal,z) \\
 &=& \int d^2\tilde\al\int \frac{d^3k}{(2\pi)^4}e^{i(\bell-R\bk_\perp)\cd\tilde\bal}e^{-ik_\pa r(z)}T_X(k,z){\cal R}(\bk) \,.
\eea
The integration of $d^2\tilde\al$  just generates a Dirac-$\delta$ function (times $(2\pi)^2$) so that
 \be\label{e:aDel}
 a^X(\bell,z) =\int \frac{d^3k}{(2\pi)^2}\de(\bell-R\bk_\perp)e^{-ik_\pa r(z)}T_X(k,z){\cal R}(\bk) \,.  \vspace{0.21cm} 
\ee
This corresponds to Eq.~(\ref{e:aXlz}) except that now, in the Dirac-$\delta$ function, $r(z)$ is replaced by $R$.
Correlating two variables $X$ and $Y$ at  redshifts $z$ and $z'$ now yields
\bea
\hspace*{-1cm}\langle  a^X(\bell,z) a^{Y *}(\bell',z')\rangle &=&\frac{1}{2\pi}\int d^3kP_{\cal R}(k)\de^2(\bell-R\bk_\perp)\de^2(\bell'-R\bk_\perp)e^{-ik_\pa (r-r')} \nonumber \\  &&  \qquad\qquad  \times T_X(k,z)T_Y^*(k,z')\\
 &=& \de^2(\bell-\bell')\frac{1}{\pi R^2}\int_0^\infty dk_\pa P_{\cal R}(k)T_X(k,z)T^*_Y(k,z')\cos(k_\pa (r-r')) \,,
\eea
where $k=\sqrt{\bell^2/R^2+k_\pa ^2}$.
The power spectrum at unequal redshift is therefore given by
\bea\label{e:Clflat2}
C^{XY}_\ell(z,z') &\simeq& \frac{1}{\pi R^2}\int_0^\infty dk_\pa P_{\cal R}(k)T_X(k,z)T_X^*(k,z')\cos(k_\pa  (r-r')) \\
&=& \frac{2\pi}{ R^2}\int_0^\infty \frac{dk_\pa}{k^3} \PP_{\cal R}(k)T_X(k,z)T_X^*(k,z')\cos(k_\pa (r-r')) \,.
\eea
This expression has two problems. First of all, the
result depends on the choice of $R$ via $k$ and via the pre-factor $1/R^2$. If $z=z'$, we can simply choose $R=r(z)$ which is the true physical distance of the flat sky.  However, for $z\neq z'$, there are different possibilities. The simplest choice is $R=\sqrt{rr'}$.
We shall see below that this is also the choice motivated by the exact expression.
The second problem is that the integrand is now rapidly oscillating and, in what concerns the numerical computation, nothing is really gained with respect to the exact calculation performed by \class. 

Therefore, let us go back and consider a term on the surface at redshift $z$, for example the density perturbation.
Writing the exponential as a sum of Legendre polynomials and spherical Bessel functions it is easy to obtain the exact standard result~\cite{Bonvin:2011bg} for two local (not integrated) variables $X$ and $Y$.

Replacing in Eq.~\eqref{e:Xbal} $\boe_z+\bal$ by $\bn$ and expanding the exponential in Legendre polynomials and spherical Bessel functions, we obtain the exact expression 
\bea\label{e:Xbn}
X(\bn,z) &=&\int \frac{d^3k}{(2\pi)^3} e^{-ir(z)\bn\cd\bk}T_X(k,z)\RR(\bk) \\
  &=& \sum_\ell i^\ell(2\ell+1)\int  \frac{d^3k}{(2\pi)^3} P_\ell(\bn\cd\hat\bk)j_\ell(kr(z))T_X(k,z)\RR(\bk) \,,
\eea
where $P_\ell$ is the Legendre polynomial of degree $\ell$. Applying the addition theorem for spherical harmonics, we find for the correlation function
\be
\langle X(\bn,z)Y(\bn',z') \rangle = \frac{1}{2\pi^2}\sum_\ell (2\ell+1)P_\ell(\bn\cd\bn')\int_0^\infty\hspace{-2mm} dkk^2j_\ell(kr(z))j_\ell(kr(z'))T_X(k,z)T^*_Y(k,z')P_\RR(k)\,, \nonumber
\ee
so that
\be
C_\ell^{X,Y}(z,z') = 4\pi\int_0^\infty \frac{dk}{k}j_\ell(kr(z))j_\ell(kr(z'))T_X(k,z)T^*_Y(k,z')\PP_\RR(k)\,. \label{e:ClXYsphere}
\ee
For the last equation we used Eq.~\eqref{e:Padim} and the fact that the correlation function is related to the power spectrum by
\be\label{e:corf-gen}
\langle X(\bn,z)Y(\bn',z')\rangle = \frac{1}{4\pi}\sum (2\ell+1)P_\ell(\bn\cd\bn')C^{X,Y}_\ell(z,z')\,.
\ee
So far, no approximation has been made and {\sc class} actually calculates the power spectra using expression Eq.~\eqref{e:ClXYsphere}.

We now use the following approximation for the Bessel functions, see \cite{Mukhanov_2004}:
\be\label{e:jMuk}
j_\ell(x) \simeq\left\{\begin{array}{ll} 0 \,,& x<L\,, \quad L=\ell+1/2 \\
\frac{\cos\left[\sqrt{x^2-L^2}-L\arccos\left(\frac{L}{x}\right) -\pi/4\right] }{\sqrt{x}(x^2-L^2)^{1/4}}\,, & x>L
\end{array}\right.
\ee
This approximation has a singularity at $x\ra L$, but is an excellent approximation for $x\gtrsim L+1$. Inserting it into Eq.~\eqref{e:ClXYsphere} we obtain
\bea\label{e:XYlessexact}
C_\ell^{XY}(z,z')&\simeq&\frac{4\pi}{r r'}\int_{\frac{\ell+1/2}{r_{\min}}}^\infty \frac{dk}{k\sqrt{k_\pa k_\pa'}} \PP_{\cal R}(k)T_X(k,z)T_Y^*(k,z')\cos\left(rk_\pa-rk_{\perp}\arccos\left(\frac{k_{\perp}}{k}\right) -\pi/4 \right)
\times \nonumber\\
&& \cos\left(r'k'_\pa-r'k'_{\perp}\arccos\left(\frac{k'_{\perp}}{k}\right) -\pi/4 \right) \\
&=& \frac{2\pi}{ r r'}\int_{\frac{\ell+1/2}{r_{\min}}}^\infty \frac{dk}{k\sqrt{k_\pa k_\pa'}}  \PP_{\cal R}(k)T_X(k,z)T_Y^*(k,z')
\Bigg\{\cos\Bigg[rk_\pa- r'k'_\pa- \nonumber\\ &&  \qquad
rk_{\perp}\left(\arccos\left(\frac{k_{\perp}}{k}\right) -\arccos\left(\frac{k'_{\perp}}{k}\right) \right) \Bigg]   \nonumber\\  &&  \qquad
+\sin\Bigg[rk_\pa+ r'k'_\pa-rk_{\perp}  \left(\arccos\left(\frac{k_{\perp}}{k}\right)+ \arccos\left(\frac{k'_{\perp}}{k}\right)\right) \Bigg]\Bigg\} \,.  \nonumber\\
&&  \label{e:XYMuk}
\eea
Here $r_{\min}=\min\{r,r'\}$ and we have defined 
\be
k_{\perp} =\frac{\ell+1/2}{r}\,, \qquad k'_{\perp} =\frac{\ell+1/2}{r'}\,, \qquad k_\pa = \sqrt{k^2-k_{\perp}^2} \,, \qquad k'_\pa = \sqrt{k^2-k_{\perp}^{\prime\;2}} \,.
\ee
 Note $rk_{\perp}=r'k'_{\perp}=\ell+1/2$, but $rk_\pa \neq r'k_\pa '$.
For $z=z'$, after replacing $\ell\ra \ell+1/2$, neglecting the rapidly oscillating sin-term and making the variable transform $kdk=k_\pa dk_\pa$, this simplifies to the flat sky approximation shown in Eq.~({\ref{e:Clflat1}}). 

However, when $z\neq z'$,  the case of interest here, the original flat sky approximation Eq.~\eqref{e:Clflat2} for $z\neq z'$ is obtained for $R^2=rr'$ only if we neglect the sin-term, set $k_\pa'=k_\pa$ and drop fully the contributions in the argument of the cos which are proportional to $rk_\perp$ which corresponds to setting $k_{\perp}=k'_{\perp}$. We have found that while keeping the sin term is not crucial,  the differences between $k_\pa'$ and $k_\pa$, as well as between  $k_{\perp}$ and $k'_{\perp}$, are.

For redshifts that are sufficiently well-separated, terms of the above form become very small. In this case the spectrum is dominated by the integrated lensing and lensing$\times$density terms which we can compute with Eqs.~({\ref{e:Cl-kd}},\ref{e:Cl-ka}), while neglecting the contributions from the local terms. However, if the redshifts are fairly close, the local terms like $D\times D$ cannot be neglected, and we must use Eq.~\eqref{e:XYMuk} to calculate them. Henceforth we refer to the collection of the terms $(C_\ell^{D},C_\ell^{\rm rsd},C_\ell^{D, {\rm rsd}})$ as the standard terms. We want to estimate their contribution for small redshift differences and also the redshift difference above which they can be neglected with respect to the lensing and lensing$\times$density contributions.

For this we examine Eq.~\eqref{e:XYMuk} in the particular case when $r$ and $r'$ are close,
\be
r'=r(1+\ep) \qquad 0<\ep\ll 1.
\ee 
Without loss of generality we assume $r<r'$ and hence $\ep>0$.
Expanding  the argument of the cos-term in Eq.~(\ref{e:XYMuk}), let us call it $a_-$, to second order in $\ep$ we find 
\bea
a_-= rk_\pa- r'k'_\pa- rk_{\perp}\left(\arccos\left(\frac{k_{\perp}}{k}\right) -\arccos\left(\frac{k'_{\perp}}{k}\right) \right)
&\simeq& -\ep rk_\pa -\frac{\ep^2}{2}\frac{k_\perp^2r}{k_\pa} +{\cal O}(\ep^3)\;.
\eea
Considering only the term $\propto\ep$ results exactly in approximation Eq.~\eqref{e:Clflat2} .
Expanding  the argument of the sin-term, let us call it $a_+$, to second order in $\ep$ we find 
\bea
a_+&=& rk_\pa+ r'k'_\pa- rk_{\perp}\left(\arccos\left(\frac{k_{\perp}}{k}\right) +\arccos\left(\frac{k'_{\perp}}{k}\right) \right)  \nonumber\\
&\simeq& 2 rk_\pa-(2\ell+1)\arccos\left(\frac{k_{\perp}}{k}\right) 
+\ep rk_\pa+\frac{\ep^2}{2}\frac{k_\perp^2r}{k_\pa} +{\cal O}(\ep^3) \,.
\eea
The sin-term oscillates with a frequency of the order of $2\ell$ which is very rapid and we can neglect it  when $k_\pa$ is not very small so that $k> k_\perp$. The cos term oscillates slowly for small $\ep$ and we should take it into account. Nevertheless, our expansion in $\ep$ cannot be trusted in the regime of very small $k_\pa$ since the $\ep^2$ terms diverge when $k_\pa\ra 0$. Furthermore, in this limit the term multiplied by  $(2\ell+1)$ in
the sin tends to $\arccos(1)=0$ and is not rapidly oscillating anymore.
In order that the approximation be deemed valid we must therefore ensure that the higher order terms in 
$\ep$ become smaller as the expansion proceeds.
The series expansion is of the general form
\bea
a_- &=& \frac{x_\pa^3}{L^2}\sum_{n=1}^\infty \al_n\left[\left(\frac{L^2}{x_\pa^2}\right)(1+{\cal O}(x_\pa^2/L^2))\ep\right]^n\\
a_+ &=& 2x_\pa +  \frac{x_\pa^3}{L^2}\sum_{n=1}^\infty \beta_n\left[\left(\frac{L^2}{x_\pa^2}\right)(1+{\cal O}(x_\pa^2/L^2))\ep\right]^n
\eea
with coefficients $|\al_n|\leq 1 $ and $|\beta_n|\leq 1$.
Here we have introduced
$x=kr$, $x'=kr' =x(1+\ep)$, $x_\perp=k_\perp r= k'_\perp r'= x'_\perp = \ell+1/2\equiv L$. Note that
\be
x'_\pa =\sqrt{x_\pa^2(1+\ep)^2 +(2\ep+\ep^2)L^2} \,.
\ee
The terms in square brackets in the series are small when $\frac{L^2}{x_\pa^2}\ep\ll 1$ but diverge for $x_\pa\ra 0$.
The approximation is therefore only valid if $x_\pa>L\sqrt{\ep}$, i.e. when these series converge.
On the other hand, even when $x_\pa=L\sqrt{\ep}$, for small $L$ the arguments become
\bea
 a_-(x_\pa=L\sqrt{\ep})  &\sim& L\sqrt{\ep^3}(1 +{\cal O}(\ep))\,,\\
 a_+(x_\pa=L\sqrt{\ep})  &\sim&  L\sqrt{\ep}(1 +{\cal O}(\ep))\,,
\eea
which may still be smaller than 1, especially for $a_-$. We can neglect the integrals only when they start oscillating (rapidly). For $x_\pa>1/\ep$ we find
\bea
 a_-(x_\pa=1/\ep)  \sim -1- \frac{L^2\ep^3}{2}(1 +{\cal O}(\ep))\quad \mbox{ while }\quad
 a_+(x_\pa=1/\ep)  \sim  \frac{2}{\ep}(1 +{\cal O}(L\ep))\,.
\eea
Hence, we choose an $x_\pa^{\max} \equiv 10\times\max\{L\sqrt{\ep},1/\ep\}$ above which both arguments of the cosine can become large and the contributions can be neglected. 

Numerical testing shows that while the second (the $\sin$) term slightly improves the overall amplitude of the approximate spectrum (the largest effect is seen for large epsilon) the integration introduces additional oscillations in $\ell$, and increases the calculation time. Therefore we neglect the second term and examine the resulting approximation.
The final expression for the approximation is:
\bea
C_\ell^{XY}(z,z')&\simeq&  \frac{2\pi}{ r^2r'}
\Bigg\{\int_0^{ x_\pa^{\max}}dx_\pa {\cal T}(k,z,z')\frac{\sqrt{x_\pa}}{x^2\sqrt{x'_\pa}}\times  \qquad\qquad \nonumber \\
 && \hspace{-1cm}\cos\left[x_\pa-x'_\pa-(\ell+1/2)\left(\arccos\left(\frac{\ell+1/2}{x}\right) -\arccos\left(\frac{\ell+1/2}{x'}\right)\right)\right]\Bigg\}\,,\qquad
\label{e:Clflat3} 
\eea
where
${\cal T}(k,z,z')  =\PP_{\cal R}( k)T_X(k,z)T_Y^*( k,z')$.

Finally, one can convert $\ep$ into the redshift difference $\delta =z'-z$. At first order $\ep$ and $\delta$ are related by
\be
r' =r(1+\ep) = r +\frac{dr}{dz}\delta= r  +\frac{1}{H(z)}\delta \qquad \mbox{ or }~ \ep=\frac{\delta}{r(z)H(z)} \,.
\ee

For  $C_\ell^{XY}(z,z')$ we have also converted the integral over $x=rk$ into an integral over $x_\pa$ using  $x_\pa dx_\pa=xdx$.
\begin{figure}[ht]
\includegraphics[width=0.95\linewidth]{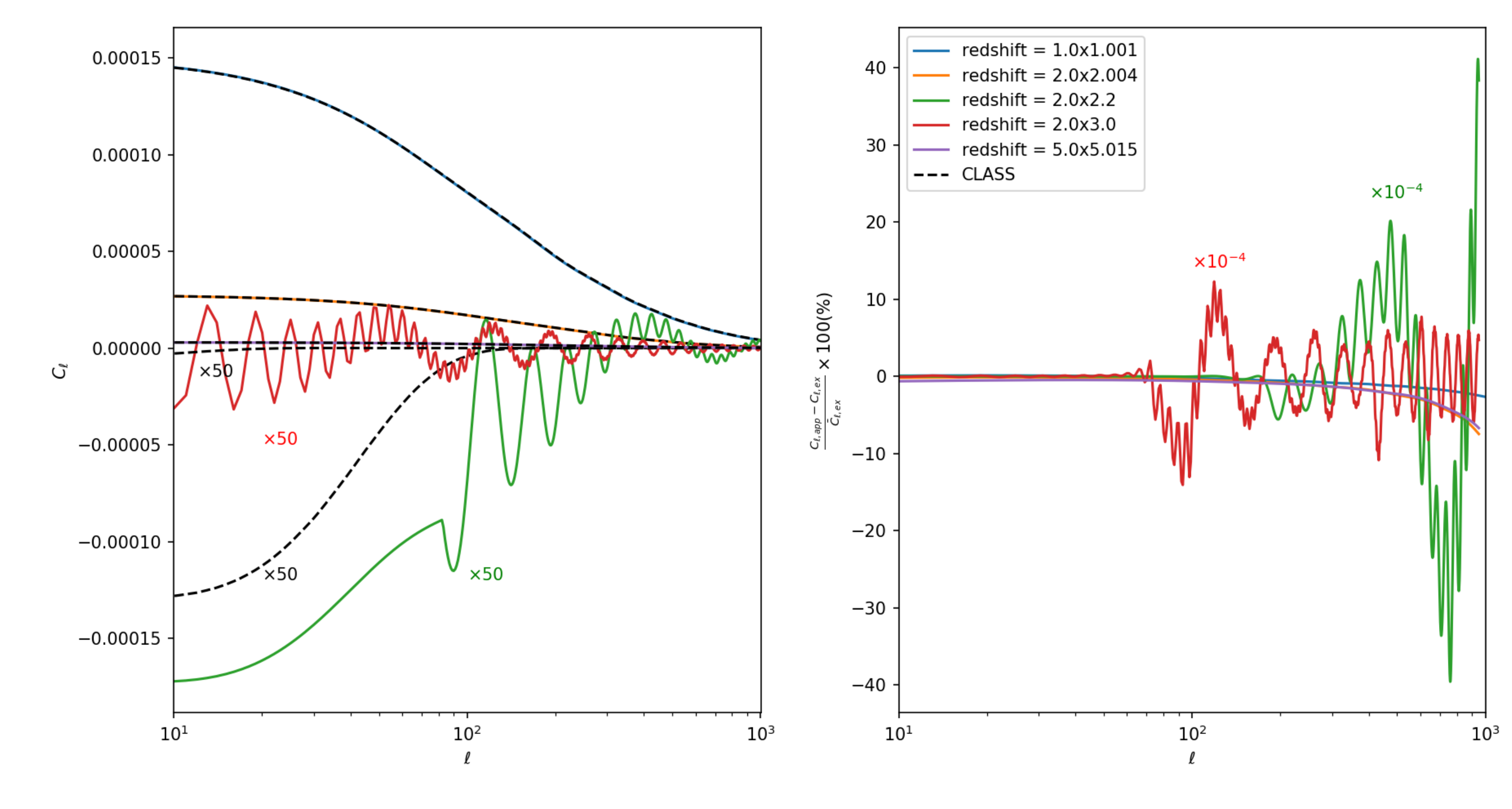}
\caption{\label{f:std-x} Left: the standard terms (density and RSD) in the flat sky approximation (solid) compared to the {\sc class} result (dashed) at redshifts  $(z,z')=(1,1.001)$, $(z,z')=(2,2.004)$, $(z,z')=(2,2.2)$, $(z,z')=(2,3)$, $(z,z')=(5,5.015)$  from top to bottom.  The 2 pairs at highest redshift separation $(z,z')=(2,2.2)$ and $(z,z')=(2,3)$ have been enhanced by a factor of $50$ to facilitate interpretation of the figure. Right: the adjusted relative differences in \%. For the redshifts $(z,z')=(2,2.2)$ and $(z,z')=(2,3)$ the error has been reduced by a factor $10^{-4}$ to make it into the plot. The true maximum error is therefore more than a factor of 1000.}
\end{figure}

In Fig.~\ref{f:std-x} we plot the  power spectrum of the standard terms   for five different redshift pairs $(z,z')$. The solid lines present our approximation Eq.~\eqref{e:Clflat3} with $X=Y=D +\dd_rV_r/\HH$. Clearly the accuracy is much worse than for equal redshifts, when the redshift difference is not very small, but note also that the amplitude is small, especially for the problematic terms with the two largest redshift separations. As we argue below, for significant redshift differences the unequal redshift terms are largely dominated by the lensing terms.
We use adjusted relative differences to determine the error. These are given by $(C_{\ell,\rm app}-C_{\ell,\rm ex})/\bar{C}_{\ell,\rm ex}$, where $C_{\ell,\rm app}$ and $C_{\ell,\rm ex}$ are respectively the power spectra from the flat sky approximation, and the {\sc  CLASS} result and the $\ell$-band average $\bar{C}_{\ell,\rm ex}$ is defined by  $\bar{C}_{\ell,\rm ex}=\sqrt{(101)^{-1}\sum_{\ell-50}^{\ell+50}(C_{\ell, \rm ex})^2}$ for 
$\ell > 50$, and for smaller $\ell$'s as many neighbouring points as are available are used. This effectively removes large spikes in the errors caused by zero-crossings of the spectra, where the relative error involves the division by a very small number of what can be a large absolute difference if the oscillations of the approximate and the exact results are out of phase.  The redshift pairs are chosen as follows: The (blue) pair  $(z,z')=(1,1.001)$ demonstrates the accuracy of the approximation in the limit tending to the equal redshift case. The pairs  $(z,z')=(2,2.004)$ and  $(z,z')=(5,5.015)$ are also chosen where the approximation is still accurate to about $5\%$. The remaining two pairs $(z,z')=(2,2.2)$ and  $(z,z')=(2,3)$ show the invalidity of the approximation when the separation in redshift increases.  However, the signal at these redshift differences is also very small, which makes this problem less relevant. In the figure the signal is enhanced by a factor of 50 for better visibility, while the error is reduced by a factor $10^4$, so that the maxima at 40\% indicate an error of a factor of 4000. Our approximation has high and low frequency oscillations in $\ell$ which are not present in the standard result. If one averages the approximation over a rather large band of $\De\ell \sim 100$, the approximation becomes better but it is still not better than the correct order of magnitude.

\begin{figure}[ht]
\includegraphics[width=1\linewidth]{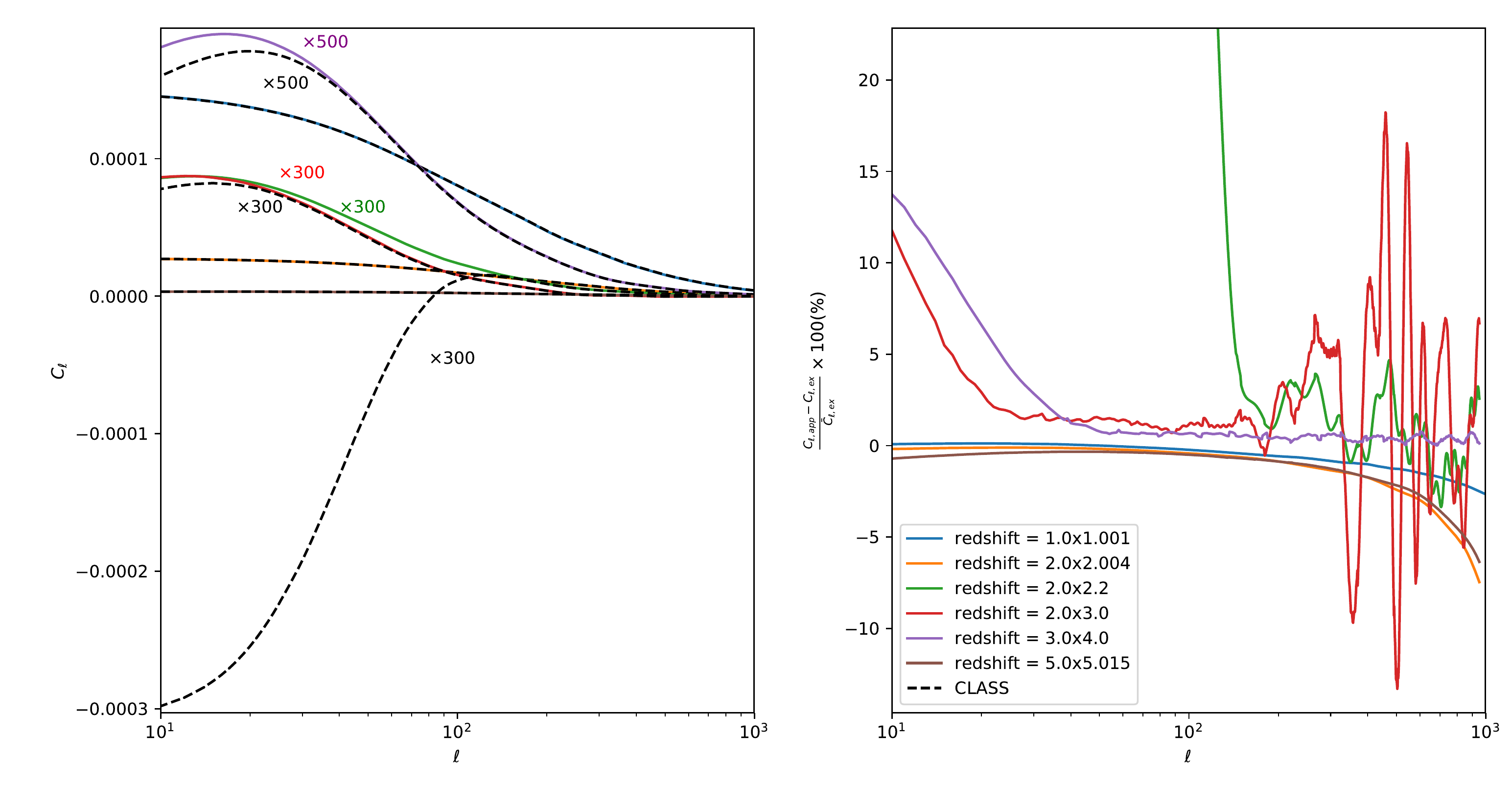}
\caption{\label{f:tot-x} Left: the total power spectrum result in the flat sky approximation (solid) compared to the {\sc class} result (dashed) at redshifts  $(z,z')=(1,1.001)$, $(z,z')=(2,2.004)$, $(z,z')=(2,2.2)$, $(z,z')=(2,3)$, $(z,z')=(5,5.015)$  from top to bottom.  The 2 pairs at highest redshift separation $(z,z')=(2,2.2)$ and $(z,z')=(2,3)$ have been enhanced by a factor of $300$ to facilitate interpretation of the figure. Right: the adjusted relative differences. }
\end{figure}

The approximation for the standard term at larger redshift separations is in general worse. However, since at large separations the lensing contribution  dominates, this problem is not very severe in the total spectra, see Fig.~\ref{f:tot-x}. 
More precisely, 
  comparing the amplitudes of the lensing terms with those of the standard terms, we find that  for redshift differences 
$$|z-z'|=\delta>\delta_1 \simeq 0.33\frac{r(z)H(z)}{1+z}~ \mbox{ and  }~\ell\sim 100 $$
the standard terms contribute less than $1\%$ to the total result. In most cases, this is then true for all $\ell\gtrsim 20$, but there are exceptions as we shall discuss.

The approximation for the standard terms at unequal redshifts is  only valid (to within $5\%$) for small redshift separations. We find that this second critical value, below which the redshift will have at most $5\%$ error, follows a power law: $\delta_0 \simeq 3.6\times10^{-4}(1+z)^{2.14}$.

\begin{figure}[ht]
\centerline{\includegraphics[width=0.55\linewidth]{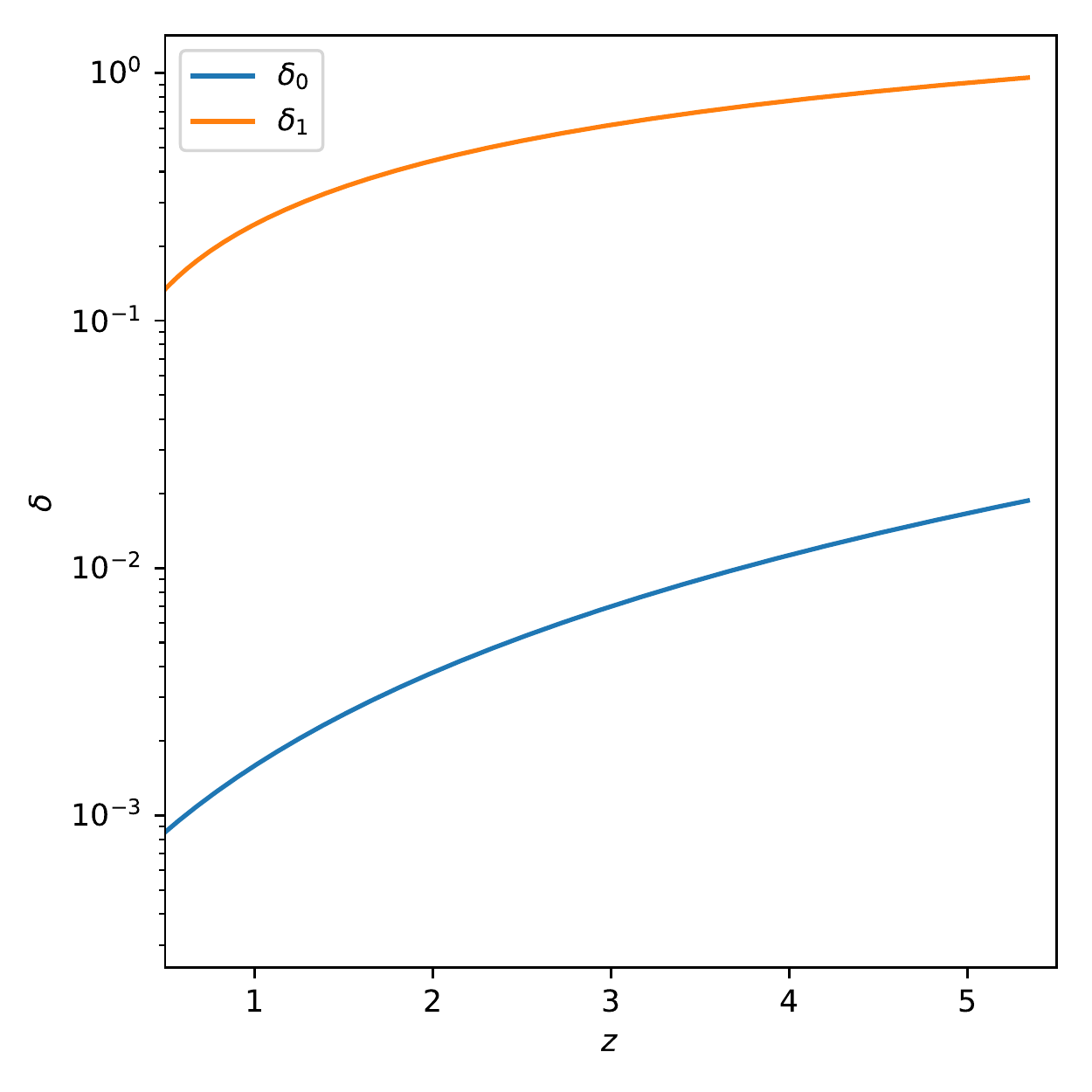}}
\caption{\label{f:Dezc} The critical redshift differences, $\delta_1\simeq0.33r(z)H(z)/(1+z)$ (orange) and $\delta_0=3.6\times10^{-4}(1+z)^{2.14}$ (blue) respectively showing the values: above which the standard terms can be neglected (for mean values between $\ell=90$ and $\ell=110$, $\bar{C}_\ell^{STD} = 1\% \bar{C}_\ell^{TOT}$) and below which the approximation is accurate to $10\%$ or less, in the unequal time correlators.
}
\end{figure}

In Fig.~\ref{f:Dezc} we show these critical separations $\delta_0$ and $\delta_1$ as functions of redshift.

For redshift differences larger than $\delta_1 $ we neglect the standard contribution, as we can again obtain a very good accuracy (1\% or better) when including only the lensing terms. 
However, our approximation for the standard terms can only be trusted for redshift differences smaller than  $\delta_0 $, where it is accurate to $5\%$ at least for $z\neq z'$ and better than 0.5\% for $z=z'$. Therefore, for redshift differences $|z-z'|\in [\delta_0,\delta_1]$ the flat sky approximation does not reach the target accuracy of 5\% and these cases need to be computed with the \class~  (or CAMB) code.

In Fig. \ref{f:tot-x} we show the total power spectrum for unequal redshifts.  Again, the adjusted relative differences are used, and the redshift pairs are chosen to illustrate: the accuracy of the approximation in the limit approaching equal redshifts with separations close to $\delta_0$  ($(z,z')=(1,1.001)$, $(z,z')=(2,2.004)$ and $(z,z')=(5,5.015)$)  and more widely separated pairs with $\delta>\delta_1$ ($(z,z')=(2,3)$ and $(z,z')=(3,4)$).  At the  redshift differences  $3-2=1$ and $4-3=1$  which are larger than $\delta_1(z=2)$ and $\delta_1(z=3)$ correspondingly, we neglect the contribution from the standard terms which is inaccurate in this regime. At sufficiently small separations (the blue, orange and purple curves), the additional contribution of the lensing terms is small but the error, mainly due to the standard term, remains below $5\%$. For large separations, on the other hand, we neglect the standard terms and include only the approximation of the remaining lensing terms.  The approximation for  $(z,z')=(3,4)$ is very good, especially above $\ell\sim 30$. However  for $(z,z')=(2,3)$ and $\ell>200$, the error actually increases to about 15\%. This is a very special redshift pair where the negative definite lensing-density correlation and the positive definite lensing-lensing correlation nearly cancel for $\ell>200$ (this can be seen  in Fig.~\ref{f:nonstd}) so that the standard terms which we neglect here contribute up to 15\%. We have checked this fact with \class{ }  which produces the same difference when neglecting the standard terms for this redshift pair. For higher redshifts the positive lensing-lensing term dominates, while for lower redshifts the negative density-lensing term dominates. In the pair $(3,4)$, above the critical separation $\delta_1(z=3) \simeq 0.5$, the lensing-lensing and lensing-density contributions for this redshift pair no longer cancel. The lensing-lensing term is truly the dominant contribution. As a result we see that, while the error increases for low $\ell$, it remains on the order of a few percent for all $\ell > 50$, and does not exhibit the particular structure that we see for the redshift pair $(z,z')=(2,3)$.
We have also tested (but not plotted) the 
pair $(2,2.2)$, still below the critical separation $\delta_1(z=2) \simeq 0.3$ where  lensing contributes more than 99\% to the signal, but much larger than $\delta_0(z=2) \simeq 0.03$. For this case the error is very large at low $\ell$ but reaches a level below $5\%$ for $\ell>130$. For lower $\ell$ the negative contribution of the standard terms to the total spectrum cannot be neglected, but is also not well modelled by our approximation. If such cases  are relevant, they have to be computed with {\sc class}. As the separation increases even further (above the critical separation), the lensing terms, which are well approximated, make up a sufficiently significant portion of the total spectrum such that the standard terms can be neglected and the results has an error below $5\%$. 
This is true also for $(z,z')=(2,2.2)$ and $\ell>130$.

Hence, there is an interplay between the accuracy of the approximation of the standard terms (which decreases for increasing separation) and their magnitude relative to the total power spectrum (which also decreases for increasing separation). If the separation is sufficiently small (e.g. blue, orange or purple curves), our approximation of the total spectra is very accurate. It degrades as the separation increases, but then it improves again once the lensing terms begin to dominate (red curve). The approximation does not meet the target accuracy for an intermediate value of the separation, $\delta_0<\delta<\delta_1$ (as an example,  $(z,z')=(2,2.2)$ while $\delta_1(z=2)\simeq 0.3$ and $\delta_0(z=2) \simeq 0.004$, we attain sub-$5\%$ error only for $\ell>130$), where our approximation of the standard terms is not accurate and we include only the lensing terms.  Finally there is the special case $(z,z')=(2,3)$ where the lensing terms nearly cancel for $\ell>200$ which degrades the approximation and induces an error of up to 15\% (green curve in Fig.~\ref{f:tot-x}).  There are also other redshift pairs at lower $z$, e.g. $(z,z')=(1,1.1)$ where the lensing terms nearly cancel, but these redshift separations are smaller and the standard terms are expected to be  relevant.
For $z>2.2$ no significant cancellation occurs anymore and the lensing term remains positive.

\subsection{Overall Performance}

Comparing Figs.~\ref{f:tot} and \ref{f:tot-x} we see that there is a large difference in the accuracy of the approximations for equal and unequal redshifts. While for equal redshifts in the range of redshifts considered, $z\in [1,5]$, the agreement of our approximation with the {\sc class} result is always better than $\sim0.5\%$, the error of the unequal redshift correlators is smaller than a few percent only for small or large  redshift differences, $\delta\leq \delta_0$ or  $\delta\geq\delta_1$. For intermediate redshift differences, $\delta_0<\delta<\delta_1$ and $\ell<100$, our approximation cannot be trusted at all. The approximation for unequal redshifts can only be used with confidence for redshifts that either are very close, $\delta\leq\delta_0$ or in the other extreme where the separation is sufficiently large, $\delta\geq \delta_1$, such that the standard terms may be safely neglected. We also show the exception to this rule, namely the redshift pair $(z,z')=(2,3)$. For $z=2$ and $z'>2.5$, the positive lensing-lensing term and the negative density-lensing term nearly cancel for $\ell>200$ so that the approximation neglecting the standard terms becomes worse again and the error increases
to nearly 15\% for some values of $\ell$.

\begin{figure}[ht]
\includegraphics[width=0.95\linewidth]{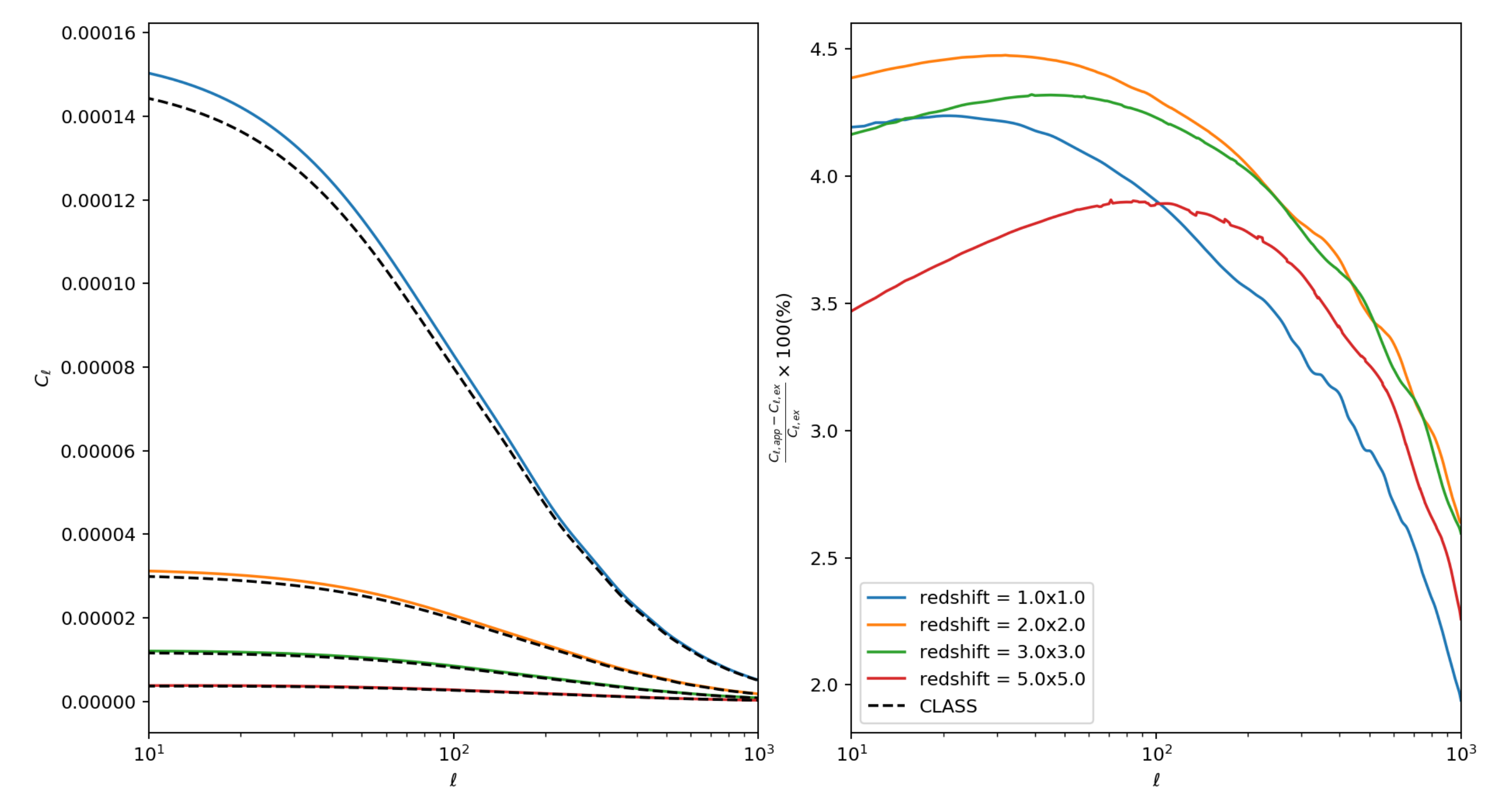}
\caption{\label{f:tot-W} Left: the windowed full power spectrum, $C_\ell^\De(z,\De z)$ in the flat sky approximation (solid) compared to the {\sc class} result (dashed) at redshifts   $z=1,~2,~3$ and $5$ from top to bottom. A top-hat window function of width $\De z = 2\De z_0$ has been applied. Right: the relative differences.} 
\end{figure}
In a real observation we cannot measure the $C_\ell(z,z')$ at exact redshifts $z,z'$ without error. First of all, even for spectroscopic surveys the redshift accuracy is finite, of order $\De z\sim 10^{-4}(1+z)$. For photometric  surveys redshift determination is much less accurate, of order $\De z\sim0.05(1+z)$ in the most optimistic case. But even if redshift accuracy is very high, in a too slim redshift bin there are few galaxies, and shot noise will prevent the determination of the $C_\ell$'s especially at high $\ell$. We therefore have also investigated windowed $C_\ell$'s defined by
\be
C_\ell(z,\De z) = \int dz_1dz_2 C_\ell(z_1,z_2)W_{\De z}(z,z_1) W_{\De z}(z,z_2)
\ee
where $W_{\De z}(z,z')$ denotes a (normalized) window function of full width $\De z$ centred at $z$. Typically one chooses a Gaussian or a tophat window. In the windowed  $C_\ell$'s unequal redshift correlators always enter and our reduced accuracy for them therefore affects
the windowed  $C_\ell$'s.
If we choose a slim window, $\De z\lesssim  2\delta_0$,  as shown in Fig.~\ref{f:tot-W}, the approximation is good with an error of 4.5\% or less for $z\leq 3$. For $z=5$, the error decreases to less than 4\%.

\begin{figure}[h!]
\includegraphics[width=0.95\linewidth,angle=0]{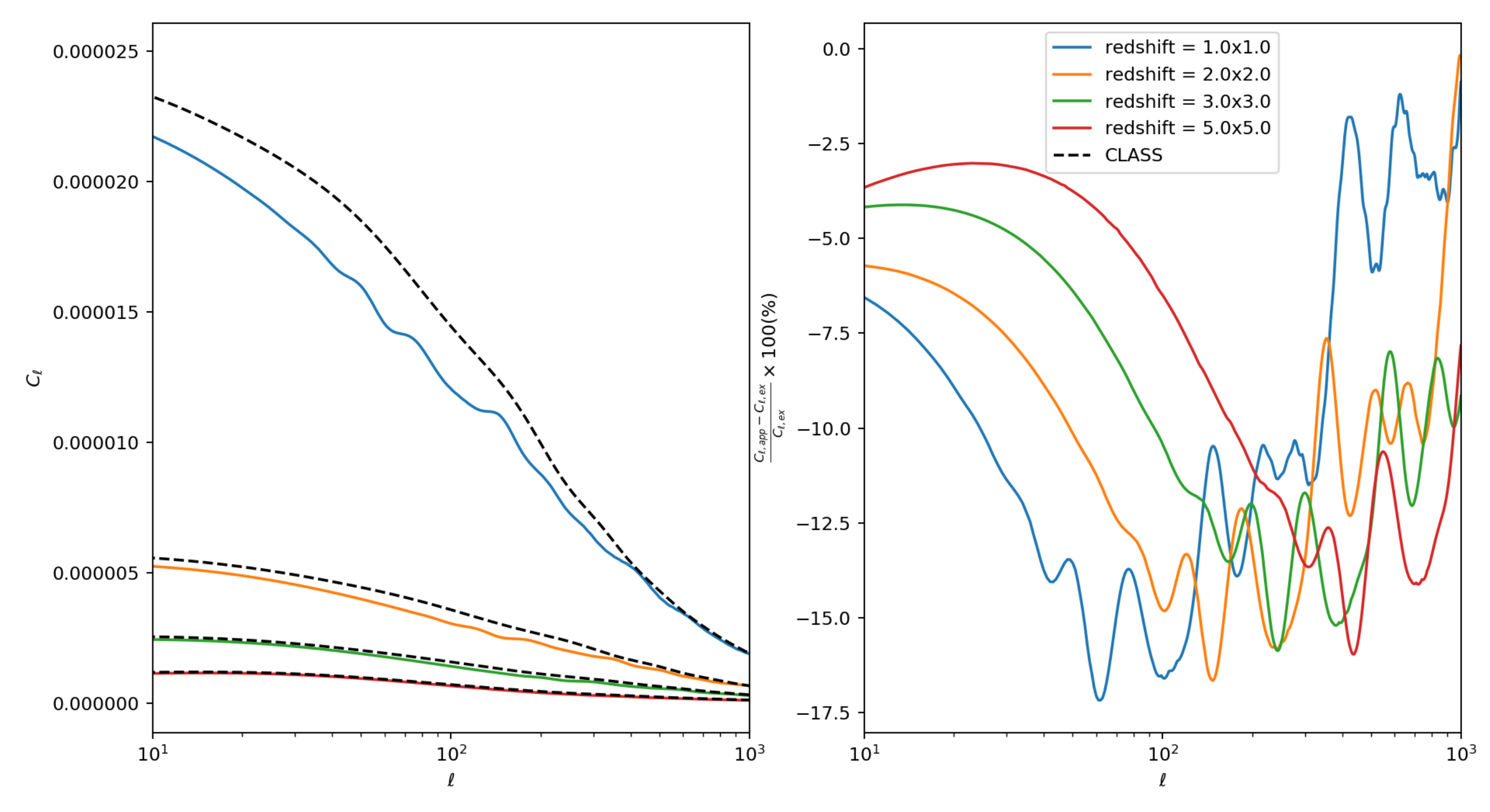}
\caption{\label{f:tot-Wphoto} Left: the windowed full power spectrum, $C_\ell^\De(z,z)$ in the flat sky approximation (solid) compared to the {\sc class} result (dashed) at redshifts   $z=1,~2,~35$ and $5$ from top to bottom. A top-hat window function of width $\De z = 0.05(1+z)$ has been applied. Right: the relative differences. Clearly the approximation is not satisfactory.} 
\end{figure}
However, if we choose a photometric window width, $\De z\gtrsim0.05(1+z)$, the accuracy degrades significantly, up to 17\%, see Fig.~\ref{f:tot-Wphoto}.

The reason for this is clear, we enter  the regime $\delta_0<\delta <\delta_1$  for which we have no good approximation. Neglecting the standard terms already for $\delta>\delta_0$ is also not a good approximation. The result then underestimated the true windowed $C_\ell$'s by nearly a factor of 10. This somewhat surprising finding shows that the standard terms do contribute significantly (about 90\% in total) also for $z\neq z'$ in the interval $\delta_0<\delta <\delta_1$. 
Also multiplying the result from the $\delta_0$-window by a factor $\delta/\delta_0$ does not give an accurate approximation. 
Therefore, a windowed correlation function cannot be determined with an accuracy better than 17\% with the flat sky approximation except for very narrow redshift bins.

\section{Comparison to the Limber Approximation}\label{s:Limber}
An approximation which is well-known, especially for lensing, is the so-called Limber approximation~\cite{Limber:1954zz,LoVerde:2008re}. We shall see that while this approximation is equivalent to the flat sky one for lensing and density-lensing correlations, it is very different and actually a bad approximation for the density and RSD terms. The fact that the Limber approximation does not work for density and RSD has  already been noted in Refs.~\cite{DiDio:2014lka,DiDio:2018unb,2020JCAP...05..010F}.

We start with the exact expression Eq.~\eqref{e:ClXYsphere} which is also used in {\sc class} to calculates the power spectra.
The Limber approximation now makes use of the fact that, for a sufficiently slowly varying  function $f(x)$ one can approximate
\be\label{e:limber1}
\int x^2f(x)j_\ell(yx)j_\ell(y'x)dx \simeq \frac{\pi}{2y^2}\de(y-y')f\left(\frac{\ell+1/2}{y}\right)
\ee
This equation is exact if $f$ is a constant.
Using it in Eq.~\eqref{e:ClXYsphere} for $X=Y=D$ yields
\bea\label{3e:den-Limber}
C^D_\ell(z,z') &=& \frac{\de(r-r')}{r(z)^2}P_{\cal R}\left(\frac{\ell+1/2}{r(z)}\right)\left|T_D\left(\frac{\ell+1/2}{r(z)},z\right)\right|^2 \\
&=& \frac{\de(z-z')H(z)}{r(z)^2}P_{\cal R}\left(\frac{\ell+1/2}{r(z)}\right)\left|T_D\left(\frac{\ell+1/2}{r(z)},z\right)\right|^2 \,.
\eea
Up to $\ell\ra \ell+1/2$, this is obtained from Eq.~\eqref{e:Clflat2} if we neglect $k_\pa$ in $k$, i.e. we set $k\simeq \ell/R$  in $P_{\cal R}(k)$ and $T_D(k,z)$ and integrate the cosine over $k_\pa $. Hence, the Limber approximation is analogous to the flat sky approximation which we used for the lensing terms.    The $\de$-function pre-factor means that for a physically sensible result we have to introduce a window function  such that
\bea
C^D_\ell(z, z', \De z) &=& \int dz_1 dz'_1 W_{\De z}(z,z_1)W_{\De z}(z',z'_1)C^D_\ell(z_1,z_1')  \nonumber \\
&=&\int \hspace{-1.2mm}dz_1  W_{\De z}(z,z_1) W_{\De z}(z',z_1) \frac{H(z_1)}{r(z_1)^2}P_{\cal R}\left(\frac{\ell+1/2}{r(z_1)}\right)T^2_D\left(\frac{\ell+1/2}{r(z_1)},z_1\right)\,.
\eea
In other words, density fluctuations are only considered correlated only if they are at equal redshift and for unequal redshifts correlations are  due entirely to overlapping window functions.

For the lensing spectra we find
\bea\label{3e:other-Limber}
C^\ka_\ell(z,z') &=& 4[(\ell+1)\ell]^2\int_0^{r_{\min}}\frac{dr}{r^2}\frac{(r(z)-r)(r(z')-r)}{r(z)r(z')r^2} P_{\cal R}\left(\frac{\ell+1/2}{r}\right) \times \nonumber \\ && \hspace{2cm} \left|T_{\Psi_W}\left(\frac{\ell+1/2}{r},z(r)\right)\right|^2 \, ,\\
C^{D,\ka}_\ell(z,z') &=& \left\{\begin{array}{c}
-2\ell(\ell+1)\frac{r(z')-r(z)}{r(z')r(z)^3} P_{\cal R}\left(\frac{\ell+1/2}{r(z)}\right)T_D\left(\frac{\ell+1/2}{r(z)},z\right)T_{\Psi_W}^*\left(\frac{\ell+1/2}{r(z)},z\right)\\
  \mbox{ if } z<z' \\
 0  \qquad \mbox{ if } z\geq z' \,, \end{array}
\right.\\
\eea
Contrary to the Limber approximation of the density, for $z=z'$  these are identical to the simplified flat sky  approximations  given in the previous section (up to $\ell\ra \ell+1/2$).

\begin{figure}[ht]
\includegraphics[width=0.9\linewidth]{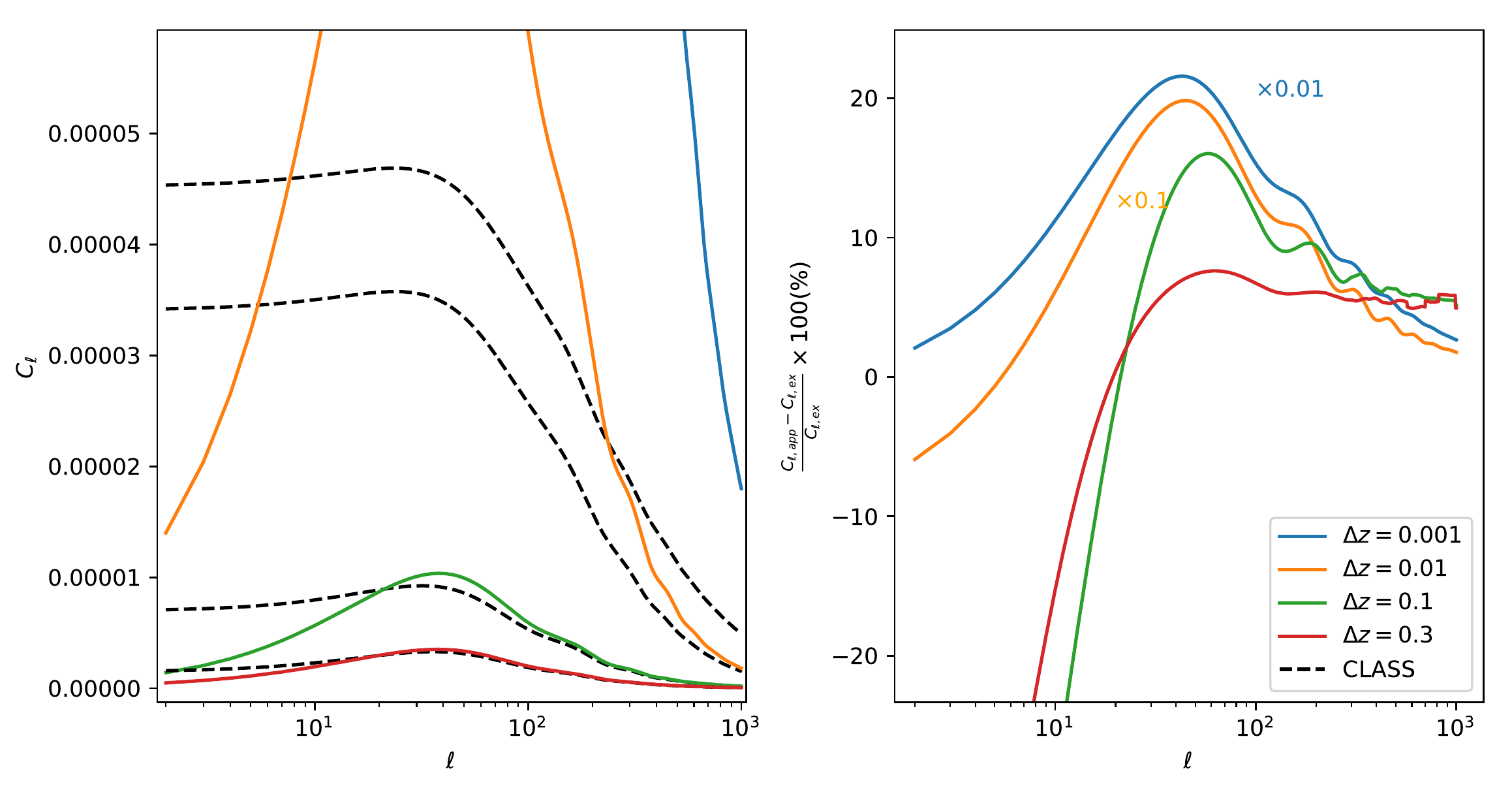}
\caption{\label{f:compD} Left: Density term at redshifts   $z=1$ in a window of width  $\De z= 10^{-3}$, $10^{-2}$,  $10^{-1}$, and $0.3$ from top to bottom. The Limber approximation (solid) and the \class~ result (black, dashed). Right: the relative differences between \class~ and the Limber approximation (solid). Note: The relative differences have been adjusted to facilitate interpretation, in particular the \% errors of the two smallest window widths (blue and orange) have been reduced by multiplication by factors 0.01 and 0.1 respectively in order to fit into the plot window. }
\end{figure}

The RSD terms are more subtle. In Appendix~\ref{A:limber-RSD} we show  that RSD terms are suppressed in the Limber approximation. It is well known that for very narrow windows, RSD can be nearly as large as the density term and therefore this approximation can at best hold for sufficiently large windows where RSD are indeed suppressed. 

In Fig.~\ref{f:compD} we compare the numerical result with the Limber approximation for the density  contribution for $z=1$ and different window widths $\De z$. Clearly, the Limber approximation cannot be trusted even at quite high $\ell$ if $\De z$ is not very large.

In Fig.~ \ref{f:compD} we see that the Limber approximation of the density term is extremely bad for slim windows, but gets increasingly more accurate as the windows widen. The accuracy is within $\sim6\%$ above $\ell=100$ for the widest window with $\Delta z = 0.3$. As expected the Limber approximation also improves with increasing $\ell$.  For $\Delta z = 0.1$, an accuracy better than  $\sim 6\%$ is achieved only above $\ell=200$.

\section{Discussion and Conclusions}\label{s:con}

In this paper we have investigated the `flat sky' approximation for galaxy number counts, which allows the calculation of the power spectra, $C_\ell^\De(z,z')$, with a simple, not heavily oscillating 1D numerical integral once the transfer function is known. For the lensing terms, the flat sky approximation is equivalent to the Limber approximation, but for density and RSD it is very different.  While the flat sky approximation of density and RSD at equal redshifts has an extremely good accuracy of below $\sim0.5\%$, the Limber approximation for the density term deviates by  $\sim5\%$ above $\ell\sim 100$ for very wide redshift bins of  $\De z\gtrsim 0.3$ for $z=1$ and by much more for slimmer redshift bins. 

The numerical calculation of the results presented is performed with a simple, unoptimised python code, and so the speed of the approximation cannot be reliably reported here. However, judging by the reduction in complexity of the various terms from their exact expressions, a speed up of around a factor $100$ may be expected from an optimised C-code compared to the full calculation by {\sc CLASS}. Recently, Ref.~\cite{jeliccizmek:2020fl} showed that the flat sky approximation, especially the use of the Limber approximation for lensing terms, can lead to a speed up of at least $\times1000$ in the calculation of the correlation function.

For equal redshifts our approximation for the standard terms, i.e. density and redshift space distortions, up to $z=3$ is accurate to within $0.3\%$ which is close to the accuracy of \class~ itself, for all values of $\ell\geq 2$. The approximation maintains excellent accuracy once the lensing terms are included, with relative error $\gtrsim 0.5\%$ up to $z=5$, for $\ell>10$. This is our first main result. 

However, for unequal redshifts, while the approximation for the lensing terms remains very good, the approximation of density and RSD degrades rapidly with growing redshift difference. For sufficiently large $\delta$, i.e. $\delta>\delta_1(\simeq 0.33r(z)H(z)/(z+1))$, the lensing terms generally dominate and the unequal redshift $C_\ell$'s can be approximated well by neglecting the standard terms. Hence, for large redshift differences we have a good approximation with errors of about 0.5\% or less for $\ell\gtrsim 60$. In addition, for sufficiently small redshift differences $\delta < \delta_0 (\simeq 3.6\times10^{-4}(1+z)^{2.14})$, the error in the standard terms is still sufficiently low that the approximaion attains $\sim 10$\% accuracy. However, for redshift differences $\delta > \delta_0$, the error of the standard term approximation becomes larger than $5\%$ and the flat sky approximation  deviates by several orders of magnitude for $\ell> 100$ once we reach $\delta_1$. Thus the approximation is not reliable for redshift differences in the interval $\delta_0 < \delta < \delta_1$. This is our second main result: for the standard terms, the flat sky approximation does not reach the target accuracy for unequal redshifts with $\delta$ in this non-negligible interval.

These results are valid for Dirac-$\de$ windows in redshift space, which is equivalent  to redshift errors of less than about $10^{-4}$, i.e. spectroscopic redshifts. The approximation  remains good, within a few percent, for windows slimmer than $\De z= 2\delta_0$. If we want to study photometric redshift bins, we have to include a window function of width $\De z= 0.05(1+z)$ or more. Doing this will always include redshift differences for which the flat sky approximation is not valid.  We have found that the approximation underestimates the true result by up to 17\%. This result is shown in Fig.~\ref{f:tot-Wphoto}. 
It is our third main result: for photometric windows, the unequal redshift standard terms are sufficiently important to degrade the the flat sky approximation considerably.

Clearly, the flat sky approximation cannot be used to estimate the $C_\ell$'s at equal redshifts with  photometric bin width. As a next step we plan to find an approximation that works also for unequal redshifts where we clearly have to go beyond both, the flat-space and the Limber approximations. This is needed to obtain a useful approximation also for photometric redshifts. 
It is also surprising that including only correlations of standard terms with redshift differences up to $\delta_0$ underestimates the windowed $C_\ell$'s by as much as a factor 10 for  bin widths of $\De z = 0.05(1+z)$. 

For spectroscopic redshifts it is often more useful to employ the correlation function since very narrow redshifts bins are plagued by shot noise on the one hand and by non-linearities in the radial direction on the other hand~\cite{Jalilvand:2019brk}. It will therefore be useful to investigate how the flat sky approximation can be used for the correlation function as calculated in Ref.~\cite{Tansella:2018sld}.

\section*{Acknowledgements}
It is a pleasure to thank Francesca Lepori for enlightening discussions. The authors acknowledge financial support from the Swiss National Science Foundation grant n$^o$ 200020-182044.

\appendix
\section{The Limber approximation for the redshift space distortion}\label{A:limber-RSD}

The exact expression for the RSD is not of the form Eq.~\eqref{e:limber1} but
\be\label{e:limber2}
\int x^2F(x)j''_\ell(yx)j''_\ell(y'x)dx \,.
\ee
More precisely, from the expressions in~\cite{Bonvin:2011bg} one finds
\bea\label{e:rsd-exact}
C^{\text{rsd}}_\ell(z_1,z_2) &=& 4\pi f(z_1)f(z_2)\int dk k^2j''_\ell(kr_1)j''_\ell(kr_2)P_\RR(k)T_D(k,z_1)T_D(k,z_2)\,,\\
C^{D, \text{rsd}}_\ell(z_1,z_2) &=& 4\pi f(z_2)\int dk k^2j_\ell(kr_1)j''_\ell(kr_2)P_\RR(k)T_D(k,z_1)T_D(k,z_2) \,.
\label{e:Drsd-exact}
\eea
Here $f(z)$ is the growth function defined in Eq.~\eqref{e:growth}.
In a $\La$CDM universe one has~\cite{Bonvin:2011bg}
\be
f(z) = 1 +\HH^{-1}\frac{\dot T_\Psi}{T_\Psi} \,.
\ee 
In order to find an approximation for integrals of the form in Eq.~\eqref{e:limber2}, we use the identity
\be\label{e:jpp-ident}
j_\ell''(x) = \frac{\ell^2-\ell-x^2}{x^2}j_\ell(x) +\frac{2}{x}j_{\ell+1}(x) \,.
\ee
To perform the required integrals we not only need a formula for integrals with equal $\ell$'s but also with $\ell$ and $\ell+1$. One might be tempted to neglect the latter, but a numerical study actually shows that both contribution to the above integral are of the same order. We therefore follow~\cite{DiDio:2018unb} and use the following crude approximation for the spherical Bessel functions to obtain integrals of unequal $\ell$
\bea
j_\ell(x) \sim  \sqrt{\frac{\pi}{2\ell+1}}\de(\ell+1/2-x) 
\eea
which yields
\bea
\frac{2}{\pi}\int_0^\infty dkk^2F(k)j_\ell(kr_1)j_{\ell+1}(kr_2)  \sim \frac{2\ell+1}{2\ell+3}F\left(\frac{\ell+1/2}{r_1}\right) \frac{\de\left(r_1\frac{2\ell+3}{2\ell+1}-r_2\right)}{r_1^2} \,.
\label{e:LmberRSD}
\eea
Inserting Eqs.~(\ref{e:LmberRSD},\ref{e:jpp-ident}) in Eq.~\eqref{e:rsd-exact} we obtain (where $r_1=r(z_1)$ and $r_2=r(z_2)$)
\bea
C^{\text{rsd}}_\ell(z_1,z_2) &\sim& \frac{4f^2(z_1)}{\ell^2}\Bigg[C^D_\ell(z_1,z_2) +C^D_{\ell+1}(z_1,z_2)  - \Bigg\{\frac{\de\left(r_1\frac{2\ell+3}{2\ell+1}-r_2\right)}{r(z)^2}P_{\cal R}\left(\frac{\ell+1/2}{r_1}\right)\times \nonumber \\  && \qquad\qquad 
T_D\left(\frac{\ell+1/2}{r_1},z_1\right)T_D\left(\frac{\ell+1/2}{r_1},z_1\right) 
+ (r_1\leftrightarrow r_2)\Bigg\}\Bigg] \,.
\eea
For the second term we have assumed that both $f(z)$ and $T_D(k,z)$ are slowly varying functions and wherein we have neglected the difference between $z_1$ and $z_2$.
We have also neglected higher order terms in $1/\ell$. 
Note that if we neglect the difference between $\ell+3/2$ and $\ell+1/2$, the second term just cancels the first term. Numerical evaluation also has shown, that for $F\simeq $ constant, the integral in Eq.~\eqref{e:LmberRSD} as a function of $r_2$ peaks even closer to $r_2=r_1$ than to $r_2= r_1\frac{2\ell+3}{2\ell+1}$. Therefore, RSD are strongly suppressed in the Limber approximation. Numerically one finds that, like for density perturbations, the Limber approximation is valid only in very wide windows where redshift space distortions are indeed strongly suppressed.

Inserting \eqref{e:jpp-ident} in Eq.~\eqref{e:Drsd-exact}, we obtain for the density-RSD correlations in the Limber approximation
\bea
C^{\text{rsd} D}_\ell(z_1,z_2) &\sim&  -\frac{2f(z_1)}{\ell}\Bigg[C^D_\ell(z_1,z_2)  - 
\frac{\de\left(r_1\frac{2\ell+3}{2\ell+1}-r_2\right)}{r(z)^2}P_{\cal R}\left(\frac{\ell+1/2}{r_1}\right)   \times  \qquad \nonumber \\  && \qquad\qquad\qquad
T_D\left(\frac{\ell+1/2}{r_1},z_1\right)T_D\left(\frac{\ell+1/2}{r_1},z_1\right) 
\Bigg] \,.
\eea
As for the pure RSD term, when neglecting the difference between $r_1$ and $r_2$, this term vanishes and therefore, for sufficiently high $\ell$, where the Limber approximation is applicable, it is negligible.

Like for the density term, the RSD Limber approximation has to be integrated over a window with some finite width $\De z$ to become physically meaningful. But for large window sizes, where the Limber approximation for the density becomes reasonably accurate, the contribution from RSD  can actually be neglected.
Also the Limber approximation of lensing--RSD is always very small. Therefore, the Limber approximation for RSD is either very poor (for slim redshift bins) or too small to be relevant. For the  wide redshift bins where the Limber approximation for the dominant density term can be sufficiently accurate, the RSD  contribution is never relevant, and neglecting it is a good approximation.

 \bibliographystyle{JHEP}
\bibliography{refs}

\end{document}